# Structure Formation Models Weaken Limits on WIMP Dark Matter from Dwarf Spheroidal Galaxies

Shin'ichiro Ando,[1,2] Alex Geringer-Sameth,[3,4] Nagisa Hiroshima,[5,6,7]
Sebastian Hoof,[8] Roberto Trotta,[3,9,10] and Matthew G. Walker[11]

[1]*GRAPPA Institute, University of Amsterdam, 1098 XH Amsterdam, The Netherlands*
[2]*Kavli Institute for the Physics and Mathematics of the Universe (Kavli IPMU, WPI), University of Tokyo, Kashiwa, Chiba 277-8583, Japan*
[3]*Department of Physics, Imperial College London, London SW7 2AZ, United Kingdom*
[4]*Department of Mathematics, Imperial College London, London SW7 2AZ, United Kingdom*
[5]*RIKEN Interdisciplinary Theoretical and Mathematical Sciences (iTHEMS), Wako, Saitama 351-0198, Japan*
[6]*Department of Physics, University of Toyama, Toyama 930-8555, Japan*
[7]*Institute of Particle and Nuclear Studies, High Energy Accelerator Research Organization (KEK), Tsukuba, Ibaraki 305-0801, Japan*
[8]*Institut für Astrophysik, Georg-August Universität, Friedrich-Hund-Platz 1, 37077 Göttingen, Germany*
[9]*Data Science Institute, William Penney Laboratory, Imperial College London, London SW7 2AZ, United Kingdom*
[10]*SISSA, Physics Department, Via Bonomea 265, 34136 Trieste, Italy*
[11]*McWilliams Center for Cosmology, Department of Physics, Carnegie Mellon University, Pittsburgh, PA 15213, United States*

Dwarf spheroidal galaxies that form in halo substructures provide stringent constraints on dark matter annihilation. Many *ultrafaint* dwarfs discovered with modern surveys contribute significantly to these constraints. At present, because of the lack of abundant stellar kinematic data for the ultrafaints, non-informative prior assumptions are usually made for the parameters of the density profiles. Based on semi-analytic models of dark matter subhalos and their connection to satellite galaxies, we present more informative and realistic *satellite priors*. We show that our satellite priors lead to constraints on the annihilation rate that are between a factor of 2 and a factor of 7 *weaker* than under non-informative priors. As a result, the thermal relic cross section can at best only be excluded (with 95% probability) for dark matter masses of $\lesssim 40\,\mathrm{GeV}$ from dwarf spheroidal data, assuming annihilation into $b\bar{b}$.

*Introduction.*—The search to uncover the nature of dark matter is one of the greatest challenges in modern physics. If dark matter is made of weakly interacting massive particles (WIMPs), as motivated by the thermal freezeout argument [1, 2] or supersymmetry [3], it can self-annihilate, producing observable gamma rays.

Dwarf spheroidal galaxies (dSphs) are associated with dark matter substructure (or subhalos). Given their proximity to us and paucity of baryons – and hence relative lack of astrophysical backgrounds – they offer the most robust environments to test the WIMP hypothesis [4–6]. In recent years, many new *ultrafaint* dSphs have been found [7]. While the small baryonic content of ultrafaint dSphs makes them promising targets for WIMP searches, the resulting dearth of stars makes it difficult to estimate their density profiles from dynamical analyses of kinematic data. A Bayesian approach can help by including additional, physical information on the parameters describing the dark matter density profile (such as a scale radius $r_s$ and a characteristic density $\rho_s$ [8]) in the form of prior probability distribution functions (PDFs). The literature to date [e.g. 9–12] has usually adopted "uninformative" *uniform priors* for both $\log r_s$ and $\log \rho_s$ (see Ref. [13] for an alternative Bayesian hierarchical analysis and Refs. [14, 15] for frequentist analyses of classical dSphs). However, such uniform priors ignore theoretical and numerical simulation results that predict the frequency distributions of subhalo parameters in the standard cold dark matter framework. While it may be appropriate to adopt such uniform priors when allowing for a variety of dark matter models, when testing WIMP dark matter specifically it is more appropriate to adopt priors derived from that model. (See Ref. [16] for a theoretical approach adopting the $r_s$-$\rho_s$ correlation expected for field halos based on a concentration-mass relation [17].) For classical dSphs with well-measured velocity dispersion profiles, the adopted priors are relatively unimportant, as the inference is dominated by the data. Therefore, we focus on the ultrafaint dSphs, where the data are sparse and a physically motivated prior becomes critical. Including structure-formation physics in the prior represents a major improvement compared to the approaches adopted to date, which use uninformative priors or are otherwise "data-driven," thus neglecting relevant physical information.

As we show in this work, subhalos occupy only specific regions of the parameter space (see the red color map in Fig. 1). Realistic constraints on WIMP annihilation, therefore, should use an informative prior distribution based on our best understanding of how dwarf galaxies form in subhalos. Such a prior is difficult to generate from $N$-body simulations, because of the limited statistics of relatively large subhalos that can host dSphs. In this *Letter*, we construct realistic *satellite priors* for the relevant parameters of the ultrafaint dSphs' dark matter distributions by using semi-analytic models based on the extended Press-Schechter (EPS) formalism combined with tidal effects on subhalo evolution, as developed in Refs. [18–20] (see also [21, 22]). We apply these novel satellite priors to obtain more realistic estimates of the gamma-ray flux from WIMP annihilation in dSphs. This results in a significant reduction of the predicted gamma-ray flux from ultrafaint dSphs compared with previous studies [9–15, 23–38].

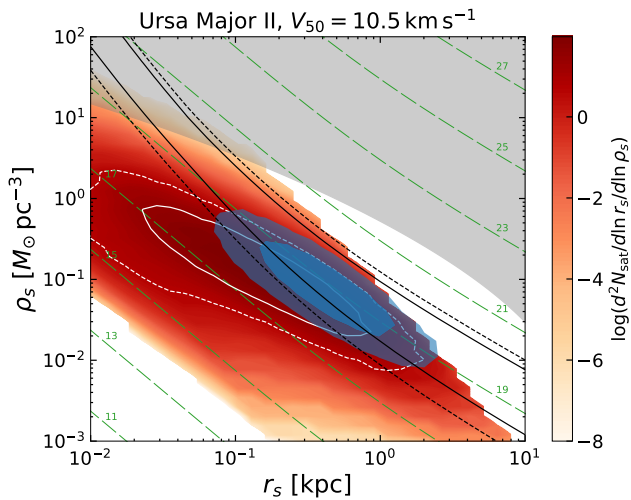

FIG. 1. Prior and posterior distributions in $(r_s, \rho_s)$ parameter space for Ursa Major II. The red color map represents the satellite number density with $V_{50} = 10.5 \,\mathrm{km\,s^{-1}}$ [cf. Eq. (2)]: $d^2 N_{\mathrm{sat}}/(d\ln r_s d\ln \rho_s)$. The open white, open black, and filled blue contours show 68% and 95% confidence/credible regions of priors, likelihood, and posteriors, respectively. The gray shaded region is the GS15 cut, excluded in previous work [10]. The green dashed curves correspond to constant values of $\log[J(0.5°)/(\mathrm{GeV^2\ cm^{-5}})]$, indicated alongside.

*Astrophysical J factor.*—The gamma-ray flux from dark matter self-annihilation from each dSph is proportional to the so-called astrophysical $J$ factor, defined as

$$J(\alpha_{\mathrm{int}}) = 2\pi \int_0^{\alpha_{\mathrm{int}}} d\psi \sin\psi \int dl \rho^2(r[l, \psi]), \quad (1)$$

where $\psi$ is the angle relative to the direction toward the center of the dSph, $\alpha_{\mathrm{int}}$ is the radius of the integration aperture, $\rho(r)$ is the dark matter density, $r^2 = l^2 + D^2 \sin^2\psi$, $l$ is line of sight distance from Earth, and $D$ is the distance to the dSph. It is commonly assumed that the density profile $\rho(r)$ is given by a spherically symmetric function, such as the Navarro-Frenk-White

(NFW) profile [8], $\rho(r) = \rho_s r_s^3/[r(r + r_s)]$, out to a tidal truncation radius $r_t$ (but see also Refs. [12, 28] for axisymmetric profiles).

*Subhalo models.*—In order to determine physically motivated priors, we adopt the semi-analytic models of subhalos developed in Refs. [19, 20]. We focus on a host halo with mass $M = 10^{12} M_\odot$ at redshift $z = 0$. The differential number of smaller halos with mass $m_a$ that accreted onto the host at redshift $z_a$ (and henceforth become subhalos), $d^2 N_{\mathrm{sh}}/(dm_a dz_a)$, is described with the EPS formalism [39], calibrated against numerical simulations [40]. After accretion, we model the evolution of the density profiles of the subhalos, which are well approximated by truncated NFW profiles [41], by taking tidal effects into account [42, 43]. This procedure predicts the distribution of subhalo variables at $z = 0$. The relevant variables for the $J$ factor are $r_s$, $\rho_s$, and $r_t$, whose joint probability density is proportional to the abundance of subhalos: $P_{\mathrm{sh}}(r_s, \rho_s, r_t) \propto d^3 N_{\mathrm{sh}}/(dr_s d\rho_s dr_t)$. In the Supplemental Material (SM), we show that the ensuing distribution of $r_s$ and $\rho_s$ is in excellent agreement with the results from numerical simulations, as is the associated subhalo mass function [19].

*Subhalo-satellite connection.*—In order to connect the subhalo population to that of the dSphs that form within them, we adopt the simple prescription given in Ref. [44]. The probability that a satellite galaxy forms in a host subhalo is given by

$$P_{\mathrm{form}}(V_{\mathrm{peak}}) = \frac{1}{2}\left[1 + \mathrm{erf}\left(\frac{V_{\mathrm{peak}} - V_{50}}{\sqrt{2}\sigma}\right)\right], \quad (2)$$

where $V_{\mathrm{peak}}$ is the peak value of the maximum circular velocity of the satellite, $V_{50}$ is where $P_{\mathrm{form}}$ is $1/2$, and we adopt $\sigma = 2.5 \,\mathrm{km\,s^{-1}}$, following Ref. [44]. (See Ref. [45] for different criteria related to reionization.)

In our model, $V_{\mathrm{peak}}$ is obtained at the time the subhalo accretes onto its host, i.e., $V_{\mathrm{peak}} = (4\pi G \rho_{s,a}/4.625)^{1/2} r_{s,a}$, where $\rho_{s,a}$ and $r_{s,a}$ are determined at accretion (see SM). According to the conventional theory of galaxy formation, we adopt a value of $V_{50}$ that allows atomic cooling to form galaxies in subhalos: $V_{50} = 18 \,\mathrm{km\,s^{-1}}$. However, Ref. [44] found that $V_{50} = 18 \,\mathrm{km\,s^{-1}}$ underpredicts the number of dSphs and their radial distribution compared with the observations, and suggested smaller values. Thus, we also adopt $V_{50} = 10.5 \,\mathrm{km\,s^{-1}}$ [44].

*Satellite prior.*—From the above distribution for subhalos we derive a distribution for satellite galaxies, which we then adopt as a prior in the analysis of kinematic data from each observed galaxy. When analyzing kinematic data, the dark matter profile of each satellite is described by parameters $(r_s, \rho_s, r_t)$. Our model results in a prior PDF:

$$P_{\mathrm{sat}}(r_s, \rho_s, r_t) \propto \frac{d^3 N_{\mathrm{sat}}}{dr_s d\rho_s dr_t} = \frac{d^3 N_{\mathrm{sh}}}{dr_s d\rho_s dr_t} P_{\mathrm{form}}(V_{\mathrm{peak}}).$$

(3)

The interpretation of this prior is that it assigns to each ultrafaint dSph an equal probability of being found in any subhalo that hosts a satellite galaxy.[1] The red color map in Fig. 1 shows the number density of satellites $d^2 N_{sat}/(d \ln r_s d \ln \rho_s)$ (after marginalization over $r_t$) for the case $V_{50} = 10.5 \, \mathrm{km \, s^{-1}}$, while the white contours show 68% and 95% highest density credible regions.

Previous studies have used uniform priors in the ($\log r_s$, $\log \rho_s$) parameter space with a sharp cut-off obtained from cosmological arguments for subhalo formation. For example, Ref. [10] uses the EPS formalism to evaluate the probability that the Milky Way hosts a subhalo with a given mass and collapse redshift and excludes parts of subhalo parameter space where this probability is low. This unphysical region is represented by the gray region in Fig. 1 and is referred to in what follows as the "GS15 cut" (see also Ref. [11]). Our model effectively allows us to replace this cut-off with a smooth transition.

*Likelihood function for observed dSphs.*—For each ultrafaint dSph, we take the data $\boldsymbol{d}$ to be summarized by the observed line-of-sight velocity dispersion $\hat{\sigma}_{los}$, the angular projected half-light radius $\hat{\theta}_h$, and distance $\hat{D}$, while the true values of these quantities are written without hats. We assume the likelihood, i.e., the probability of obtaining data $\boldsymbol{d}$ for a dSph given model parameters $\boldsymbol{\theta}$, $P(\boldsymbol{d}|\boldsymbol{\theta}) \equiv \mathcal{L}(\boldsymbol{\theta})$, to be

$$\mathcal{L}(\boldsymbol{\theta}) = \prod_{x \in \{\theta_h, \sigma_{los}, D\}} \frac{1}{\sqrt{2\pi\sigma_x^2}} \exp\left[-\frac{(\hat{x}-x)^2}{2\sigma_x^2}\right], \quad (4)$$

where $\sigma_x$ is the measurement uncertainty on $\hat{x}$ and $x$ is the model value. For classical dSphs, velocity dispersion profiles provide additional important information but for the sparsely observed ultrafaints there is little to be gained in using more than the single value $\hat{\sigma}_{los}$. Our data are detailed in the SM.

According to the virial theorem, for a spherical system in dynamic equilibrium, the line-of-sight velocity dispersion is given by [51]

$$\sigma_{los}^2 = \frac{4\pi G}{3} \int_0^\infty dr \, r\nu_\star(r)M(r), \quad (5)$$

where $M(r)$ is the enclosed mass within radius $r$ (assumed dominated by dark matter) and $\nu_\star(r)$ is the stellar density profile, for which we adopt a Plummer sphere: $\nu_\star(r) = 3[1 + (r/R_h)^2]^{-5/2}/(4\pi R_h^3)$ with $R_h = D\theta_h$. For a subhalo characterized by $\boldsymbol{\theta} = (r_s, \rho_s, r_t, \theta_h, D)$, we compute $\sigma_{los}$ with Eq. (5) and use this in Eq. (4) to evaluate the likelihood.

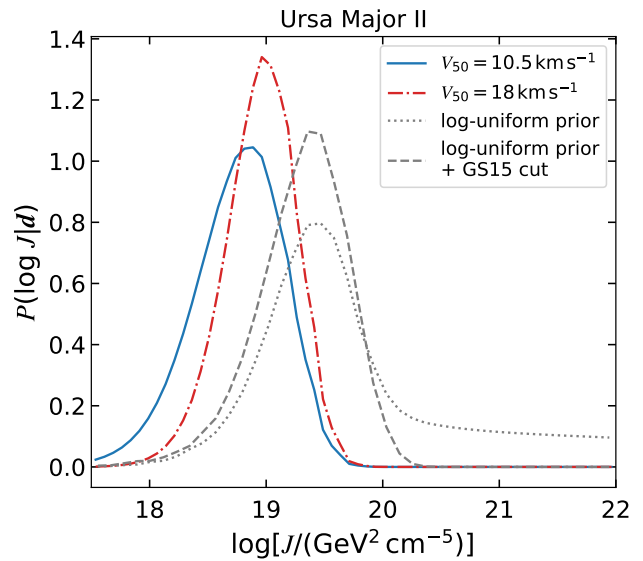

FIG. 2. Posterior distributions of $J(0.5°)$ for Ursa Major II, obtained with log-uniform priors for $r_s$ and $\rho_s$ (dotted), log-uniform priors with the GS15 cut that corresponds to the gray shaded region in Fig. 1 (dashed), satellite priors with $V_{50} = 18 \, \mathrm{km \, s^{-1}}$ (dot-dashed), and $V_{50} = 10.5 \, \mathrm{km \, s^{-1}}$ (solid). Median and $1\sigma$ credible regions for $\log[J/(\mathrm{GeV^2 \, cm^{-5}})]$ from these posteriors are $19.58^{+1.54}_{-0.51}$, $19.34^{+0.34}_{-0.41}$, $18.96^{+0.28}_{-0.32}$, and $18.78^{+0.35}_{-0.42}$, respectively.

*Posterior distribution.*—Applying Bayes' theorem using the satellite prior [Eq. (3)] and the likelihood [Eq. (4)], we obtain the posterior PDF of the subhalo quantities $\boldsymbol{\theta}$ as $P(\boldsymbol{\theta}|\boldsymbol{d}) \propto P_{sat}(\boldsymbol{\theta})\mathcal{L}(\boldsymbol{\theta})$. In Fig. 1, we show the marginal posterior PDF on $r_s$ and $\rho_s$ for Ursa Major II as filled blue contours encompassing 68% and 95% probability (highest probability density regions), while the likelihood (maximized over $\theta_h$ and $r_t$, assuming $r_t \gg R_h$) is shown by the black contours. For log-uniform priors on $r_s$ and $\rho_s$, the posterior would trace these iso-likelihood contours. The degeneracy between $r_s$ and $\rho_s$, which occurs for ultrafaint dwarfs, can be broken by the additional information supplied by the prior.

Given values for $r_s$, $\rho_s$, and $r_t$, as well as distance $D$ for each dSph, we evaluate $J(0.5°)$ [30]. In Fig. 1, we show contours of constant $\log[J/(\mathrm{GeV^2 \, cm^{-5}})]$ for Ursa Major II, from which one can see the impact of adopting different priors (log in this work is base-10). We ignore substructure boosts of the annihilation rate for dSphs, as they are at most a few tens of percent (see SM). In addition, we ignore the errors on $D$ (i.e. set $\sigma_D = 0$). The fractional errors in distance are much smaller than those on $\sigma_{los}$ and we have checked that ignoring them has no impact on our results. The priors on $\theta_h$ and $\sigma_{los}$ are uniform over positive values.

In Fig. 2, we show the marginalized posterior distribution on $\log J(0.5°)$ of Ursa Major II for four different priors. Compared with a log-uniform prior, the satel-lite prior moves the posterior toward the bottom-left corner in $(r_s, \rho_s)$ parameter space, which yields systematically smaller $J$ factors. In the case of $V_{50} = 10.5\,\text{km}\,\text{s}^{-1}$ ($18\,\text{km}\,\text{s}^{-1}$), the median of the $J$ distributions with the satellite prior is 3.6 (2.4) times smaller than when using a log-uniform prior with GS15 cut. We discuss all the other ultrafaints in the SM.

---

[1] The framework could be extended by adopting stellar-mass–halo-mass relations [46–49] as an additional factor in the likelihood. But these relations are known to have large uncertainty for faint galaxies [50], and we choose not to include them.

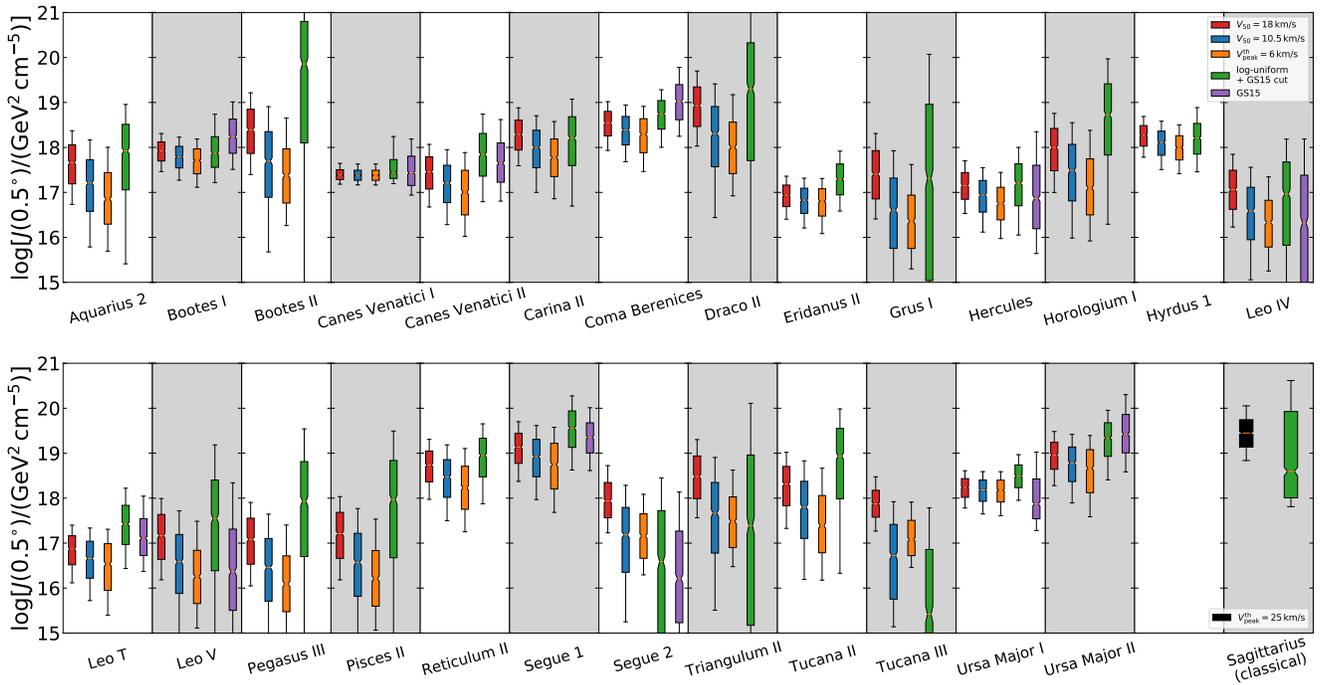

FIG. 3. Box-whisker diagram showing median values and equal-tailed 68% (boxes) and 95% (whiskers) credible intervals of the $J(0.5°)$ posteriors. For comparison, posteriors with priors uniform in $(\log r_s, \log \rho_s)$ with the GS15 cut and previous results from Ref. [10] are shown. The classical dSph Sagittarius uses a different satellite formation threshold of $V_{\text{peak}}^{\text{th}} = 25\,\text{km}\,\text{s}^{-1}$ [45] (see SM for the other classical dSphs).

Figure 3 summarizes the median values and 68% and 95% credible intervals for $J(0.5°)$ for all ultrafaint dSphs (and the classical dSph Sagittarius, which we find has one of the largest $J$ factor but has been considered in very few previous studies). The figure compares the results when using the satellite prior with different assumptions for $V_{\text{peak}}$ to the results using log-uniform priors with the GS15 cut. We find that for ultrafaint dSphs exhibiting the largest $J$ factors in previous analyses, adopting the satellite prior produces $J$ distributions whose medians are systematically smaller. This generic result also holds true in comparison with earlier work [10, 11, 33]. In Fig. 3, we also show the $J$ factors resulting from replacing the probability of a satellite forming in a host subhalo, Eq. (2), by a step function $P_{\text{form}} = \Theta(V_{\text{peak}} - V_{\text{peak}}^{\text{th}})$ with $V_{\text{peak}}^{\text{th}} = 6\,\text{km}\,\text{s}^{-1}$. Although it may seem implausible that such small subhalos host galaxies in strong radiation fields after cosmic reionization, such a scenario has been suggested from the observed numbers and distribution of satellites [44]. In this extreme case, we observe that the

$J$ factor distributions shift even further toward smaller values.

*Constraints on WIMP annihilation.*—We now quantify the impact of the satellite priors on annihilation cross section limits using Fermi-LAT gamma-ray data. We use a sample of 31 dSphs, adding Boötes II, Segue 2, Triangulum II, and Tucana III to the 27 dSphs in Ref. [35]. We do not include Sagittarius due to its proximity to the Galactic plane. Our data selection, background modeling and sampling techniques are as in Ref. [35]: we use around 11 years of Pass 8 (R3) data [52] in conjunction with the 4FGL source catalogue [53].

For the classical dSphs,[2] we use the $J$ factor posteriors of Ref. [33] as priors in our analysis since $J$ factors of classical dSphs are well-constrained by the large number of member stars and are relatively insensitive to the choice of prior distribution. For the ultrafaint dwarfs, we compute marginal $J$ factor distributions under the following priors:

(i) Uniform prior on $(\log r_s, \log \rho_s)$ with the GS15 cut;

(ii) Satellite prior, Eq. (3), $V_{50} = 18\,\text{km}\,\text{s}^{-1}$;

(iii) Satellite prior, Eq. (3), $V_{50} = 10.5\,\text{km}\,\text{s}^{-1}$;

---

[2] These are Carina, Draco, Fornax, Leo I, Leo II, Sculptor, Sextans, and Ursa Minor.

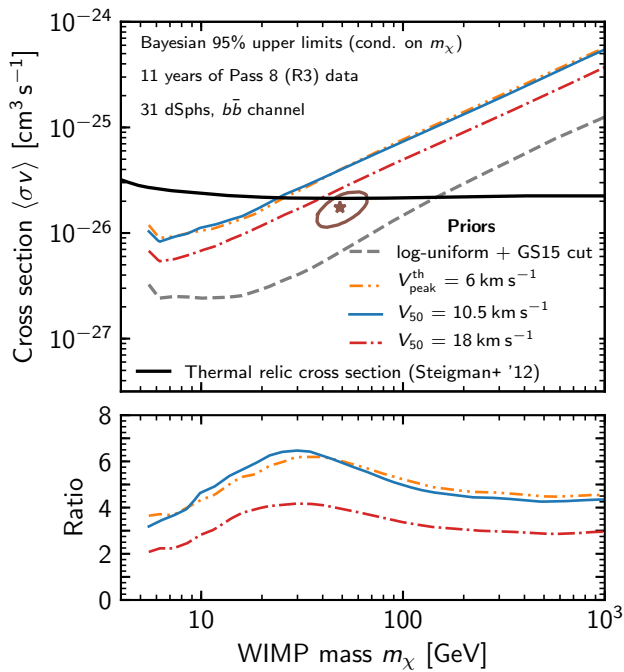

FIG. 4. Limits on the WIMP annihilation cross section $\langle \sigma v \rangle$ ($b\bar{b}$ channel) for different prior choices. *Top:* Upper limits at 95% credibility (conditioned on the WIMP mass $m_\chi$). The star and surrounding region indicate the parameter point and $2\sigma$ confidence levels associated with a possible Galactic centre excess [55] (see also [56–61]), respectively. *Bottom:* Ratios of the cross-section upper limits obtained with satellite priors and those with log-uniform prior with GS15 cut; i.e., how much *weaker* the limits derived from the satellite priors are.

(iv) Satellite prior, Eq. (3), step function replacing Eq. (2), $V_{\text{peak}}^{\text{th}} = 6 \, \text{km s}^{-1}$.

We implement these $J$ distributions as priors in the gamma-ray analysis. As described in Ref. [35], we use the T-Walk algorithm [54] to compute the full posterior over the 64-dimensional parameter space of dark matter mass, $m_\chi$, and annihilation cross section, $\langle \sigma v \rangle$, along with the $J$ factors and diffuse background normalization parameters of each dSph.

Figure 4 (top) compares the resulting upper limits on the cross section under the different prior assumptions. Limits on $\langle \sigma v \rangle$ are obtained from the posterior distribution conditioned on WIMP mass annihilating to a $b\bar{b}$ final state (in the SM, we also show limits for the $\tau^+\tau^-$ channel). Figure 4 (bottom) shows ratios normalized to the limit obtained from the prior (i) above. Satellite priors result in limits that are *weaker* by a factor of between $\sim 2$ and $\sim 7$ than uninformative priors. In particular, under informative priors the thermal relic cross section can only be excluded with 95% probability for $m_\chi \lesssim 40 \, \text{GeV}$ at best (and $m_\chi \lesssim 25 \, \text{GeV}$ at worst), in contrast to $m_\chi \lesssim 150 \, \text{GeV}$ for uninformative priors.

*Conclusions.*—In this *Letter*, we introduced satellite priors based on physical modeling of dark matter subhalos and a semi-analytical formalism connecting them to the Milky Way's population of satellite galaxies. Our informative priors assign a higher probability to regions of $(\log r_s, \log \rho_s)$ parameter space where subhalos and satellites tend to be found, in contrast to the uniform priors in $(\log r_s, \log \rho_s)$ space widely adopted in the literature. Our priors therefore better reflect the physical mechanisms of subhalo and satellite formation in the cold dark matter picture. When applying our informative satellite priors to the analysis of 11 years of Fermi-LAT data from 31 dSphs, we found that the limits on dark matter annihilation cross section are substantially weaker (between a factor of 2 and 7) compared to using the less informative log-uniform priors. This is a consequence of a systematic shift of most of the $J$ factors to smaller values induced by the informative prior, which downweighs the parameter space region where dSphs are unlikely to form. We conclude that physically motivated priors for the properties of dSphs, which encompass as much as possible our understanding of structure and galaxy formation, are crucial for interpreting the particle properties of dark matter.

This work was supported by Royal Society International Exchanges Scheme between Imperial College London (RT) and University of Amsterdam (SA), and JSPS/MEXT KAKENHI Grant Numbers JP17H04836, JP18H04578 (SA), and JP18H04340 (SA and NH). AGS and RT are supported by Grant ST/N000838/1 from the Science and Technology Facilities Council (UK). RT was partially supported by a Marie-Skłodowska-Curie RISE (H2020-MSCA-RISE-2015-691164) Grant provided by the European Commission. SH gratefully acknowledges funding by the Alexander von Humboldt Foundation and the German Federal Ministry of Education and Research. We acknowledge the UK Materials and Molecular Modelling Hub, which is partially funded by EPSRC (EP/P020194/1), and the Imperial College Research Computing Service (doi:10.14469/hpc/2232) for providing computational resources. AGS and MGW acknowledge support from NASA's Fermi Guest Investigator Program, Cycle 9, grant NNX16AR33G.

# Supplemental Material: Structure Formation Models Weaken Limits on WIMP Dark Matter from Dwarf Spheroidal Galaxies

Shin'ichiro Ando,[1, 2] Alex Geringer-Sameth,[3, 4] Nagisa Hiroshima,[5, 6, 7]
Sebastian Hoof,[8] Roberto Trotta,[3, 9, 10] and Matthew G. Walker[11]

[1] *GRAPPA Institute, University of Amsterdam, 1098 XH Amsterdam, The Netherlands*
[2] *Kavli Institute for the Physics and Mathematics of the Universe (Kavli IPMU,
WPI), University of Tokyo, Kashiwa, Chiba 277-8583, Japan*
[3] *Department of Physics, Imperial College London, London SW7 2AZ, United Kingdom*
[4] *Department of Mathematics, Imperial College London, London SW7 2AZ, United Kingdom*
[5] *RIKEN Interdisciplinary Theoretical and Mathematical Sciences (iTHEMS), Wako, Saitama 351-0198, Japan*
[6] *Department of Physics, University of Toyama, Toyama 930-8555, Japan*
[7] *Institute of Particle and Nuclear Studies, High Energy Accelerator
Research Organization (KEK), Tsukuba, Ibaraki 305-0801, Japan*
[8] *Institut für Astrophysik, Georg-August Universität,
Friedrich-Hund-Platz 1, 37077 Göttingen, Germany*
[9] *Data Science Institute, William Penney Laboratory,
Imperial College London, London SW7 2AZ, United Kingdom*
[10] *SISSA, Physics Department, Via Bonomea 265, 34136 Trieste, Italy*
[11] *McWilliams Center for Cosmology, Department of Physics,
Carnegie Mellon University, Pittsburgh, PA 15213, United States*
(Dated: February 26, 2020; revised 14 July 2020)

# I. ANALYTIC MODELS OF SUBHALOS

We summarize the analytic models of dark matter subhalos on which the satellite priors are based (see Refs. [1, 2] for full details).

## A. Subhalo evolution

In order to derive physically motivated priors for dSphs, we adopt the analytic models of subhalo accretion and tidal evolution developed by Ref. [1], which are in very good agreement with the results of numerical simulations of subhalo properties such as their mass function. We focus on a host halo with mass $M = 10^{12} M_\odot$ at redshift $z = 0$. The differential number of smaller halos with mass at accretion $m_a$, which accreted onto the host at redshift $z_a$ (thus becoming subhalos), $d^2 N_{\rm sh}/(dm_a dz_a)$, is described with the extended Press-Schechter formalism [3] that is well calibrated and tuned compared with results of numerical simulations [4].

The virial radius $r_{200}$ of the halo just accreted onto its host is obtained via $m_a = 4\pi [200 \rho_c(z_a)] r_{200,a}^3/3$, where $\rho_c(z_a)$ is the critical density at $z_a$. We then obtain the scale radius via $r_{s,a} = r_{200,a}/c_a$, where $c_a$ is the concentration parameter, drawn from a log-normal distribution with mean $\bar{c}_{200}(m_a, z_a)$ [5] and standard deviation $\sigma_{\log c} = 0.13$ [6]. We finally obtain the halo characteristic density at accretion time, $\rho_{s,a}$, through $m_a = 4\pi \rho_{s,a} r_{s,a}^3 f(c_a)$, where $f(x) = \ln(1+x) - x/(1+x)$.

After accretion, a subhalo characterized by the variables $(m_a, z_a, c_a)$ starts losing its mass through the tidal effect exerted by the host's gravitational field. We model the mass-loss rate as

$$\frac{dm}{dt} = -A \frac{m(z)}{\tau_{\rm dyn}(z)} \left[ \frac{m(z)}{M(z)} \right]^\zeta,$$  (1)

where $\tau_{\rm dyn}(z)$ is the dynamical timescale [7] and $M(z)$ is the host mass [5] at redshift $z$. Through simple Monte Carlo simulations of the mass loss process through tidal force and in comparison with the results of $N$-body simulations, we find that Eq. (1) holds for a large dynamic range of $m/M$, where both $A$ and $\zeta$ parameters in Eq. (1) only weakly depend on $M$ and $z$ [1]. We solve Eq. (1) from $z = z_a$ to 0 with an initial condition of $m(z_a) = m_a$, to obtain the subhalo mass at $z = 0$ after tidal stripping, denoted by $m_0$.

Numerical simulations indicate that, after tidal effects are accounted for, the density profile of the subhalos will remain nearly NFW up to a sharp truncation radius $r_{t,0}$ [8]. We obtain the scale radius $r_{s,0}$ and the characteristic density $\rho_{s,0}$ both at the present time after the tidal stripping, by following Ref. [9] that parameterized both $r_{s,0}/r_{s,a}$ and $\rho_{s,0}/\rho_{s,a}$ as a function of the mass ratio $m_0/m_a$. Then, by solving the enclosed mass condition, $m_0 = 4\pi \rho_{s,0} r_{s,0}^3 f(r_{t,0}/r_{s,0})$, we obtain $r_{t,0}$. This way, we obtain all the subhalo parameters after tidal evolution, $(m_0, \rho_{s,0}, r_{s,0}, r_{t,0})$, given mass $m_a$, redshift $z_a$, and the concentration $c_a$ at the time of the subhalo's accretion.

## B. Generating the list of subhalos

We subdivide the 3-dimensional $(\log m_a, z_a, \log c_a)$ parameter space with an equally-spaced grid along all dimensions. For each point $i$ in the grid, we compute the differential number of subhalos that form at that point in parameter space [5, 6], and associate with it a weight $w_i$ proportional to the number of subhalos forming in a finite cube around that point:

$$w_i \propto \left( \frac{d^2 N_{\rm sh}}{dm_a dz_a} \right)_{m_{a,i}, z_{a,i}} (\Delta m_a)_i (\Delta z_a)_i P(c_{a,i}|m_{a,i}, z_{a,i})(\Delta c_a)_i,$$  (2)

where $d^2 N_{\rm sh}/dm_a dz_a$ is the subhalo accretion rate [4], $P(c_a|m_a, z_a)$ is the log-normal distribution for $c_{200}$ [5, 6], and $(\Delta x)_i$ is the $i$th bin width for the quantity $x$. The weights $w_i$ are normalized to the total number of subhalos ever accreted onto the host,

$$\sum_i w_i = N_{\rm sh,total} \equiv \int dm_a \int dz_a \frac{d^2 N_{\rm sh}}{dm_a dz_a}.$$  (3)

The weights $w_i$ thus indicate how many subhalos with parameters indexed by $i$ exist today in a Milky Way-like halo.

By following the procedure above, we can compute all the quantities after the tidal evolution, $(m_0, \rho_{s,0}, r_{s,0}, r_{t,0})$, for each entry $i$. For example, the distribution of $m_0$ is the subhalo mass function, which is in excellent agreement with results of numerical simulations for various sets of host mass and redshift (see Fig. 2 of Ref. [1]).

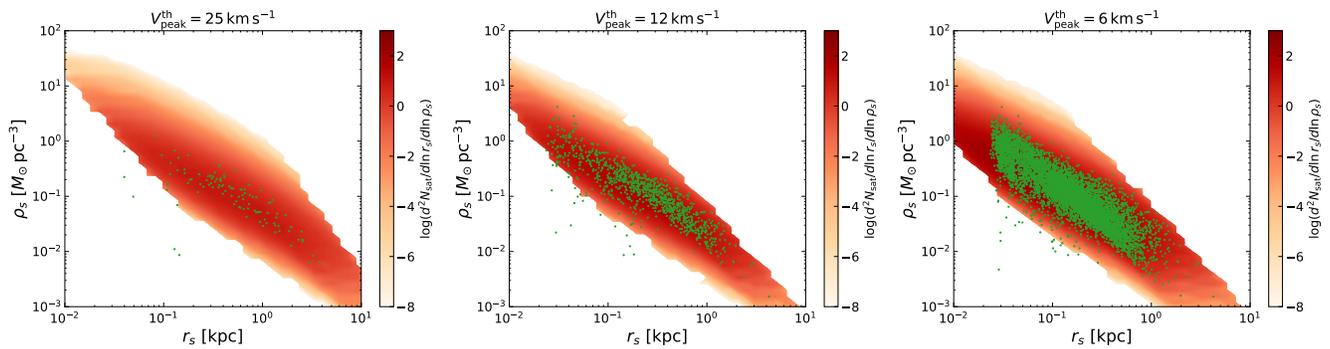

FIG. 1. Satellite density in $(\log r_s, \log \rho_s)$ space from our model, $d^2 N_{\rm sh}/(d\ln r_s d\ln \rho_s)$, for three different thresholds for forming satellites in the subhalos, $V^{\rm th}_{\rm peak} = 25$, 12, and 6 km s$^{-1}$. The green dots show subhalos found in the numerical simulation Via Lactea II [10] obeying the same threshold criteria on $V_{\rm peak}$.

In Fig. 1, we show the subhalo number density in the $(\log r_s, \log \rho_s)$ parameter space (here and in the following, we omit the subscript 0), for three different threshold values of $V_{\rm peak}$, the peak value of the subhalo's maximum circular velocity (which occurs at accretion in our model): $V^{\rm th}_{\rm peak} = 25$ km s$^{-1}$ (corresponding to subhalos hosting classical dSphs), $V^{\rm th}_{\rm peak} = 12$ km s$^{-1}$, and $V^{\rm th}_{\rm peak} = 6$ km s$^{-1}$ (describing two possibilities for ultrafaint dSphs). Our prior distribution for satellites is proportional to the subhalo number density multiplied by formation probability of a satellite in the given subhalo $P_{\rm form}$, which is a function of $V_{\rm peak}$ (see main text). The satellite prior distribution in the $(\log r_s, \log \rho_s)$ parameter space shown in Fig. 1 is then obtained after marginalizing over $r_t$. For this figure we have adopted a step function formation probability of $P_{\rm form}(V_{\rm peak}) = \Theta(V_{\rm peak} - V^{\rm th}_{\rm peak})$ to facilitate comparison with numerical simulations.

Figure 1 also shows the values of $(\log r_s, \log \rho_s)$ of each subhalo found in the N-body simulations Via Lactea II (VL-2) [10] (green points). The density profile data for the subhalos of VL-2 are given in terms of $V_{\rm max}$ and $r_{\rm max}$, which we convert to $r_s$ and $\rho_s$ assuming an NFW profile. In all the cases shown in Fig. 1, we see good agreement between the analytic models adopted in this work and VL-2 simulations.

The simulation contains some subhalos at small $r_s$ and small $\rho_s$, in regions of vanishing small prior density. These subhalos might have an anomalous merging history, or not be fully virialized, and hence are not captured by our model. However, we note that if the prior were to include them this would shift the resulting $J$ factors to even lower values.

## II. SUB-SUBHALO BOOSTS OF THE DWARF SPHEROIDALS

Dwarf galaxies form in subhalos, and they might host their own subhalos, i.e., sub-subhalos. Since the dark matter annihilation rate is boosted in the presence of such sub-subhalos, we need to assess the importance of this effect, i.e. the annihilation boost factor [2] of the dSphs. Previous work estimated the effect and found it to be negligibly small [11, 12]. Here we revisit the question in the context of the improved subhalo model presented in the previous section.

The gamma-ray emissivity profile from WIMP annihilation in the sub-subhalos traces the radial distribution of the sub-subhalos, for which we adopt $[1 + (r/r_s)^2]^{-3/2}$, while that of the smooth component follows the NFW profile squared. The subhalo hosting a dSphs suffers from tidal stripping down to radius $r_t$, after which the luminosity from the sub-subhalos $L_{\rm sh}$ and from the smooth component $L_{\rm sm}$ (within dSphs' virial radius at accretion, $r_{{\rm vir},a}$), will change to $L_{\rm sh}(< r_t)$ and $L_{\rm sm}(< r_t)$, respectively, as follows:

$$\frac{L_{\rm sh}}{L_{{\rm sh},a}} = \frac{r_s^3 \left[ \sinh^{-1}(r_t/r_s) - r_t/\sqrt{r_t^2 + r_s^2} \right]}{r_{{\rm vir},a}^3 \left[ \sinh^{-1}(r_{{\rm vir},a}/r_{s,a}) - r_{{\rm vir},a}/\sqrt{r_{{\rm vir},a}^2 + r_{s,a}^2} \right]}, \tag{4}$$

$$\frac{L_{\rm sm}}{L_{{\rm sm},a}} = \frac{\rho_s^2 r_s^3 \left[ 1 - (1 + r_t/r_s)^{-3} \right]}{\rho_{s,a}^2 r_{s,a}^3 \left[ 1 - (1 + r_{{\rm vir},a}/r_{s,a})^{-3} \right]}, \tag{5}$$

where the expressions in the right-hand side of Eqs. (4) and (5) can be obtained from the volume integral of $[1 +$

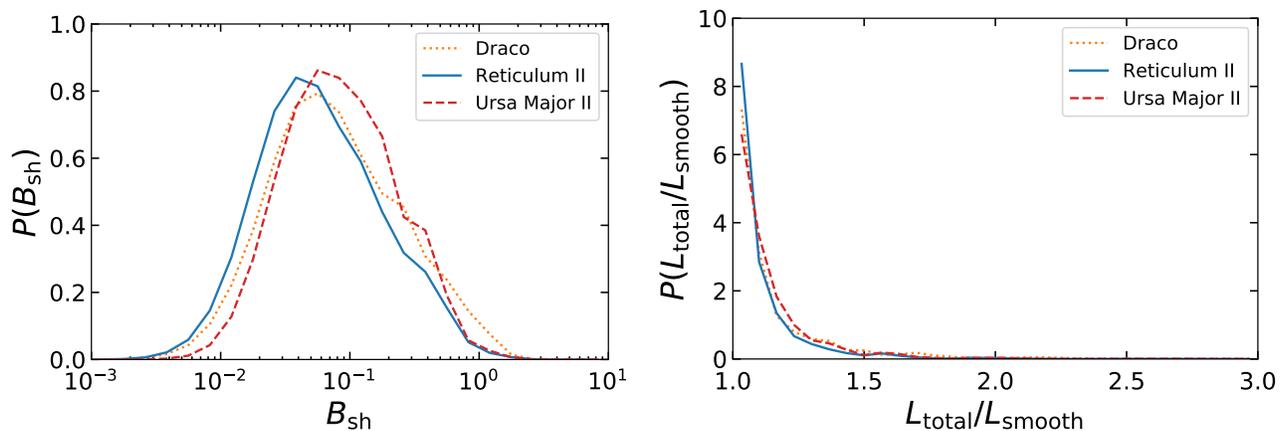

FIG. 2. Posterior probability distributions of the subhalo boost factor, $B_{\rm sh}$, and the ratio between the total luminosity and that from the smooth NFW component in the absence of subhalos, $L_{\rm total}/L_{\rm smooth} = 1 - f_{\rm sh}^2 + B_{\rm sh}$. The cases of Draco, Reticulum II, and Ursa Major II are shown, with satellite forming conditions of $V_{\rm peak}^{\rm th} = 25$ km s$^{-1}$ (Draco) and $V_{50} = 10.5$ km s$^{-1}$ (Reticulum II and Ursa Major II).

$(r/r_s)^2]^{-3/2}$ and $\rho_{\rm NFW}^2(r)$. The annihilation boost factor after stripping, $B_{\rm sh} \equiv L_{\rm sh}/L_{\rm sm}$, is obtained from

$$B_{\rm sh} = \left( \frac{L_{\rm sh}/L_{{\rm sh},a}}{L_{\rm sm}/L_{{\rm sm},a}} \right) B_{{\rm sh},a}(m_a, z_a), \tag{6}$$

where $B_{{\rm sh},a}(m_a, z_a)$ is the boost factor at accretion, which is fully characterized by the mass $m_a$ and redshift $z_a$ at accretion; cf. Fig. 5 in Ref. [2]. Likewise, the sub-subhalo mass fraction $f_{\rm sh}$ is corrected as

$$f_{\rm sh} = f_{{\rm sh},a} \frac{L_{\rm sh}/L_{{\rm sh},a}}{\rho_s r_s^3 f(r_t/r_s)/[\rho_{s,a} r_{s,a}^3 f(r_{\rm vir,a}/r_{s,a})]}. \tag{7}$$

The total luminosity of a dSph, $L_{\rm total}$, is then related to that in the absence of subhalos, $L_{\rm smooth}$, by (Eq. 15 of Ref. [2])

$$L_{\rm total} = \left(1 - f_{\rm sh}^2 + B_{\rm sh}\right) L_{\rm smooth}. \tag{8}$$

In Fig. 2, we show the posterior probability distributions of $B_{\rm sh}$ and $L_{\rm total}/L_{\rm smooth}$ obtained from our model for a few dwarfs: Draco (with $V_{\rm peak}^{\rm th} = 25$ km s$^{-1}$), Reticulum II and Ursa Major II (with $V_{50} = 10.5$ km s$^{-1}$). The boost factor $B_{\rm sh}$ is much smaller than one, in contrast with the results of Ref [1, 2], which found $B_{\rm sh} \sim 1$ for large range of host masses at $z = 0$. This is due to tidal effect (Eq. (6)) that strips the regions outside of the tidal radius $r_t$ away. As a result, the enhancement of the annihilation rate is very minor, at most a few tens of percent, and this does not depend much on which dSphs we are interested in. Hence we can safely ignore the subhalo boost to the annihilation rate in the dSphs.

## III. OBSERVATIONAL DATA

We require three measured properties for each dSph: heliocentric distance $\hat{D}$, angular projected half-light radius $\hat{\theta}_h$, and line-of-sight velocity dispersion $\hat{\sigma}_{\rm los}$. Our likelihood function is the product of Gaussians for each of these quantities, i.e. $\mathcal{L}(x) = (2\pi\sigma_x^2)^{-1/2} \exp[-(\hat{x} - x)^2/(2\sigma_x^2)]$, where $x$ is one of $D$, $\theta_h$, or $\sigma_{\rm los}$. If a measured quantity is given as $\hat{x}_{-b}^{+a}$ then we take $\sigma_x = (a+b)/2$. For some dSphs the line of sight velocity dispersion is unresolved and only upper limits on $\sigma_{\rm los}$ are reported. In this case we determine $\hat{\sigma}_{\rm los}$ and $\sigma_{\sigma_{\rm los}}$ by finding the Gaussian likelihood that would have produced the reported upper limits. For example, Ref. [13] finds a posterior $1\sigma$ upper limit of 1.4 km/s and a $2\sigma$ upper limit of 2.6 km/s for the velocity dispersion of Segue 2. We imagine the posterior as arising from a 1d Gaussian likelihood with a uniform prior over all non-negative values of $\sigma_{\rm los}$. Then we find the mean and standard deviation of the Gaussian likelihood that yields the same $1\sigma$ and $2\sigma$ upper limits as stated above. In this case we obtain $\hat{\sigma}_{\rm los} = 0.53$ km/s and $\sigma_{\sigma_{\rm los}} = 1.11$ km/s.

Our dynamical analysis models each dSph as a spherical system whose stellar distribution follows a Plummer profile. We therefore use the "circularized" half-light radius, defined as the geometric mean of the semimajor and semiminor axes of the 2d elliptical isophote which contains half the total flux. Some studies report measurements of the circularized half-light radius directly while others give the semimajor axis of the half-light ellipse $\theta_{h,\mathrm{maj}}$ along with the ellipticity $e$. These quantities are related by $\theta_h = \sqrt{1-e}\,\theta_{h,\mathrm{maj}}$. In order to propagate uncertainties we generate values according to approximate error distributions for $\theta_{h,\mathrm{maj}}$ and $e$ as follows. If a measured quantity (i.e. $\theta_{h,\mathrm{maj}}$ or $e$) is reported as $\hat{x}_{-b}^{+a}$ we generate samples $x_i$ according a "split normal distribution": $P(x_i) = \mathcal{N}(x_i \mid \hat{x}, a)$ if $x_i > \hat{x}$ and $P(x_i) = \mathcal{N}(x_i \mid \hat{x}, b)$ if $x_i < \hat{x}$, where $\mathcal{N}(x_i \mid \mu, \sigma) = (2\pi\sigma^2)^{-1/2}\exp[-(x_i - \mu)^2/(2\sigma^2)]$ is the normal probability density. We generate independent samples of $\theta_{h,\mathrm{maj}}$ and $e$, create the resulting distribution of $\theta_h$ samples, and find the 15.9, 50, and 84.1 percentiles in order to compute $\hat{\theta}_h$ and the upper and lower $1\sigma$ error bars. Where possible we adopt $\hat{\theta}_h$ determined by fits of Plummer surface brightness profiles to the photometric data since we assume Plummer profiles in the virial theorem. However, for some dSphs half-light radii derived from fits to exponential or King surface brightness profiles are the only ones available. Even so, the functional form of the stellar density profile $\nu_*(r)$ only enters the velocity dispersion prediction as part of the integrand in the virial theorem. Therefore, the velocity dispersion is not particularly sensitive to the shape of $\nu_*(r)$ but rather to the location where $r\nu_*(r)$ peaks, which generally occurs around the half-light radius for any functional form.

Tables I and II summarize the observational data for the ultrafaint and classical dSphs. For most dSphs we use the Plummer profile fits in Ref. [14], which is the latest study to perform a uniform analysis over the large majority of dSphs. For distances we adopt those in Ref. [14] which are compiled from other studies (and without uncertainties, though these are negligible for our purposes in any case). We use the velocity dispersions compiled in Ref. [15] when available.

## IV. PRIORS AND POSTERIORS FOR DWARF GALAXIES

### A. Classical dwarfs

Figure 3 shows the prior, the likelihood, and the posterior (68% and 95% credible or confidence regions) on $r_s$ and $\rho_s$ for all the classical dwarfs, where the satellite priors are obtained with $V_{\mathrm{peak}}^{\mathrm{th}} = 25$ km s$^{-1}$ following Ref. [44]. For the classical dSphs, one can see a good degree of overlap between priors and likelihoods, leading to a posterior that is, in most cases, only slightly different from what would be obtained from the likelihood and the GS15 cut together. The one exception is Sagittarius, where the posterior is noticeably more constrained than the likelihood, due to the degeneracy between $\rho_s$ and $r_s$.

The resulting $J$ factors (integrated within $0.5°$) are summarized in Fig. 4 and Table III. We generally observe a good agreement between our central values using log-uniform priors with a GS15 cut (green) and the results of GS15 [45] (purple), which used a full Jeans analysis of velocity data for individual stars instead of summary statistics (i.e. global velocity dispersion). It is our use of summary statistics that makes the uncertainty in our results with log-uniform priors larger than in GS15. Comparing to the results when adopting the satellite priors (black), we observe as expected a reduction in the uncertainty with respect to the case of log-uniform priors, while maintaining the good agreement with the full analysis of GS15.

TABLE I. Measured properties of ultrafaint dSphs: distance, half-light radius $\hat{\theta}_h$, and line-of-sight velocity dispersion $\hat{\sigma}_{\rm los}$. Distances and half-light radii without references are from Tables 1 and 3 of Ref. [14]. Half-light radii are derived from fits to Plummer profiles unless otherwise indicated. Dwarfs with unresolved velocity dispersions have $\hat{\sigma}_{\rm los} \pm \sigma_{\hat{\sigma}_{\rm los}}$ values derived from the posterior quantiles given in footnotes (see Sec. III).

| Name | Distance [kpc] | | $\hat{\theta}_h$ [arcmin] | | $\hat{\sigma}_{\rm los}$ [km s$^{-1}$] | |
|------|-----|-----|-----|-----|-----|-----|
| Aquarius 2 | $107.9 \pm 3.3$ | [16] | $3.96^{+0.71}_{-0.67}$ | [16] | $5.4^{+3.4}_{-0.9}$ | [16] |
| Bootes I | 66 | | $8.34^{+0.29}_{-0.29}$ | | $4.6^{+0.8}_{-0.6}$ | [17] |
| Bootes II | 42 | | $2.73^{+0.43}_{-0.41}$ | | $10.5 \pm 7.4$ | [18] |
| Canes Venatici I | 218 | | $5.33^{+0.21}_{-0.21}$ | | $7.6 \pm 0.4$ | [19] |
| Canes Venatici II | 160 | | $1.16^{+0.23}_{-0.22}$ | | $4.6 \pm 1.0$ | [19] |
| Carina II | $36.2 \pm 0.6$ | [20] | $7.04^{+0.73}_{-0.70}$ | [20] | $3.4^{+1.2}_{-0.8}$ | [21] |
| Coma Berenices | 44 | | $4.47^{+0.30}_{-0.29}$ | | $4.6 \pm 0.8$ | [19] |
| Draco II | 20 $\pm 3$ | [22] | $2.27^{+0.95}_{-0.79}$ | [22]a | $2.9 \pm 2.1$ | [23] |
| Eridanus II | 380 | | $1.42^{+0.15}_{-0.15}$ | | $6.9^{+1.2}_{-0.9}$ | [24] |
| Grus I | 120 | | $0.53^{+0.56}_{-0.49}$ | | $-4.65 \pm 6.28$ | [25]b |
| Hercules | 132 | | $3.13^{+0.30}_{-0.29}$ | | $3.7 \pm 0.9$ | [26] |
| Horologium I | 79 | | $1.34^{+0.30}_{-0.9}$ | | $4.9^{+2.8}_{-0.9}$ | [27] |
| Hyrdus 1 | $27.6 \pm 0.5$ | [28] | $6.64^{+0.46}_{-0.43}$ | [28]a | $2.7^{+0.5}_{-0.4}$ | [28] |
| Leo IV | 154 | | $2.30^{+0.28}_{-0.27}$ | | $3.3 \pm 1.7$ | [19] |
| Leo T | 417 | | $1.10^{+0.13}_{-0.13}$ | | $7.5 \pm 1.6$ | [19] |
| Leo V | 178 | | $0.72^{+0.30}_{-0.27}$ | | $3.7^{+2.3}_{-1.4}$ | [29] |
| Pegasus III | 215 | | $0.66^{+0.24}_{-0.21}$ | [30] | $5.4^{+3.0}_{-2.5}$ | [30] |
| Pisces II | 182 | | $0.90^{+0.15}_{-0.14}$ | | $5.4^{+3.6}_{-2.4}$ | [31] |
| Reticulum II | 30 | | $3.58^{+0.15}_{-0.15}$ | | $3.6^{+1.0}_{-0.7}$ | [32] |
| Segue 1 | 23 | | $2.95^{+0.42}_{-0.40}$ | | $3.9 \pm 0.8$ | [33] |
| Segue 2 | 35 | | $3.31^{+0.29}_{-0.29}$ | | $0.53 \pm 1.11$ | [13]c |
| Triangulum II | 30 | | $1.43^{+0.44}_{-0.41}$ | | $-3.64 \pm 3.13$ | [34]d |
| Tucana II | 57 | | $9.83^{+1.66}_{-1.11}$ | [35]a | $8.6^{+4.4}_{-2.7}$ | [25] |
| Tucana III | 25 $\pm 2$ | [36] | $6.00^{+0.80}_{-0.60}$ | [36] | $-0.62 \pm 0.93$ | [37]e |
| Ursa Major I | 97 | | $5.32^{+0.30}_{-0.29}$ | | $7.6 \pm 1.0$ | [19] |
| Ursa Major II | 32 | | $9.15^{+0.46}_{-0.45}$ | | $6.7 \pm 1.4$ | [19] |

a Half-light radius derived from fit to exponential profile
b Based on the 16th and 50th percentiles of the $\hat{\sigma}_{\rm los}$ posterior
c Based on the 1$\sigma$ and 2$\sigma$ upper limits from the $\hat{\sigma}_{\rm los}$ posterior
d Based on the 90% and 95% upper limits from the $\hat{\sigma}_{\rm los}$ posterior
e Based on the 90% and 95.5% upper limits from the $\hat{\sigma}_{\rm los}$ posterior

TABLE II. The same as Table I but for classical dSphs.

| Name | Distance [kpc] | $\hat{\theta}_h$ [arcmin] | $\hat{\sigma}_{\rm los}$ [km s$^{-1}$] |
|------|-----|-----|-----|
| Carina | 105 | $8.08^{+0.10}_{-0.10}$ | $6.6 \pm 1.2$ [38] |
| Draco | 76 | $8.15^{+0.10}_{-0.09}$ | $9.1 \pm 1.2$ [39, 40] |
| Fornax | 147 | $16.51^{+0.13}_{-0.13}$ | $11.7 \pm 0.9$ [38] |
| Leo I | 254 | $3.05^{+0.03}_{-0.03}$ | $9.2 \pm 0.4$ [41] |
| Leo II | 233 | $2.43^{+0.03}_{-0.03}$ | $7.4 \pm 0.4$ [42] |
| Sagittarius | 26 | $205.10^{+9.25}_{-9.12}$ [15]a | $11.4 \pm 0.7$ [43] |
| Sculptor | 86 | $9.14^{+0.08}_{-0.08}$ | $9.2 \pm 1.1$ [38] |
| Sextans | 86 | $13.80^{+0.13}_{-0.13}$ | $7.9 \pm 1.3$ [38] |
| Ursa Minor | 76 | $12.28^{+0.15}_{-0.16}$ | $9.5 \pm 1.2$ [39] |

a Half-light radius derived from fit to King profile

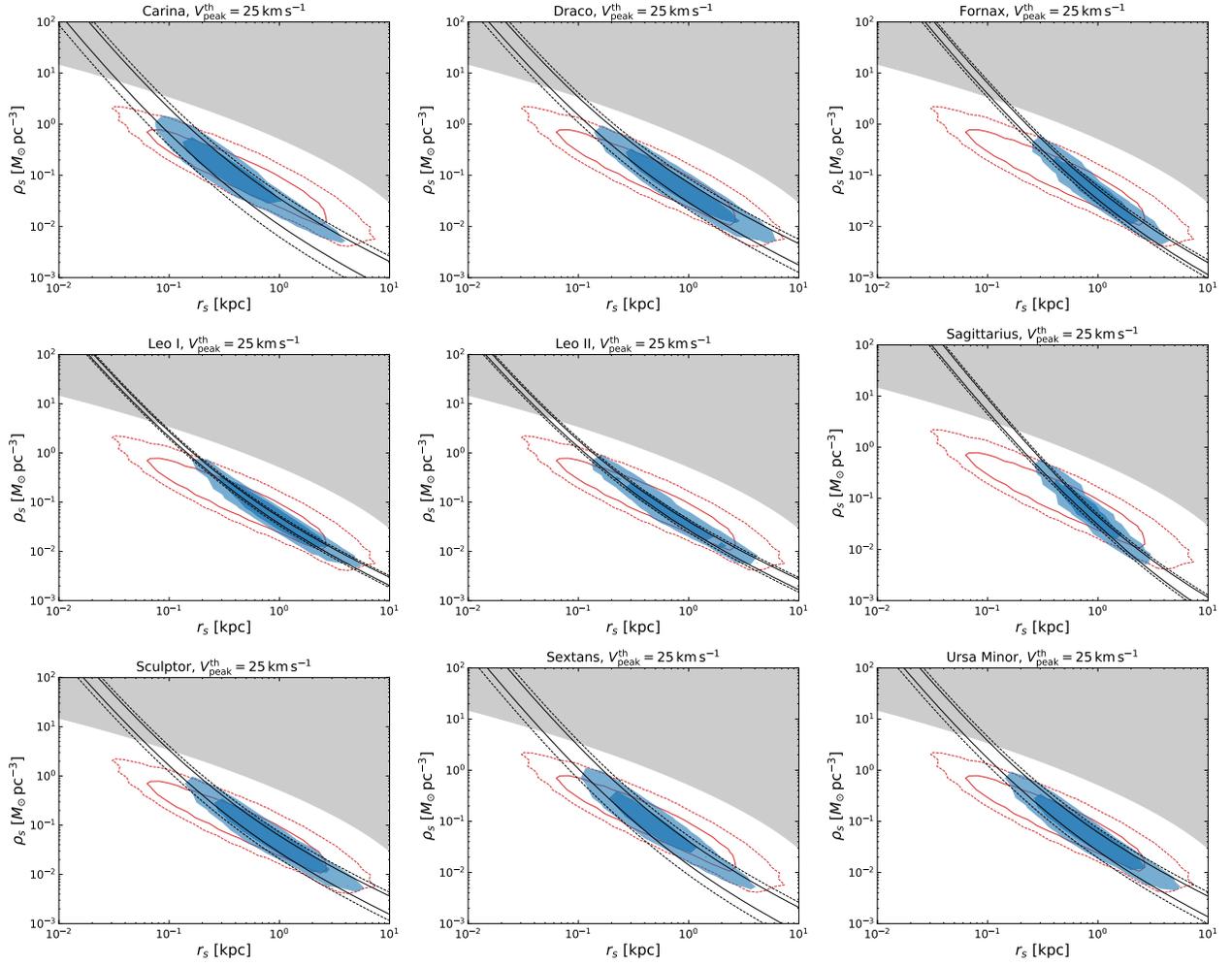

FIG. 3. Satellite prior (red), likelihood (black), and posterior (filled blue) contours in the $(r_s, \rho_s)$ plane for classical dSphs, showing 68% and 95% credible regions (prior and posterior) and 68%, 95% confidence regions (likelihood). The prior is based on the satellite forming condition of $V_{\rm peak}^{\rm th} = 25$ km s$^{-1}$. The gray shaded region shows the cosmology cut implemented in Ref. [45].

TABLE III. The medians and 68% credible intervals (defined by the 16th and 84th percentiles of the posterior distribution) for $\log[J(0.5^\circ)/({\rm GeV}^2\,{\rm cm}^{-5})]$ for the classical dSphs.

| Name | Log-uniform + GS15 cut | $V_{\rm peak}^{\rm th} = 25$ km s$^{-1}$ |
|---|---|---|
| Carina | $17.92^{+0.39}_{-0.38}$ | $18.00^{+0.23}_{-0.25}$ |
| Draco | $18.85^{+0.26}_{-0.26}$ | $18.75^{+0.20}_{-0.22}$ |
| Fornax | $18.02^{+0.56}_{-0.23}$ | $18.19^{+0.19}_{-0.17}$ |
| Leo I | $17.89^{+0.18}_{-0.16}$ | $17.74^{+0.09}_{-0.08}$ |
| Leo II | $17.77^{+0.21}_{-0.20}$ | $17.59^{+0.11}_{-0.10}$ |
| Sagittarius | $18.59^{+1.33}_{-0.59}$ | $19.45^{+0.30}_{-0.31}$ |
| Sculptor | $18.65^{+0.27}_{-0.25}$ | $18.59^{+0.18}_{-0.19}$ |
| Sextans | $18.16^{+0.47}_{-0.37}$ | $18.31^{+0.22}_{-0.23}$ |
| Ursa Minor | $18.68^{+0.34}_{-0.27}$ | $18.67^{+0.19}_{-0.20}$ |

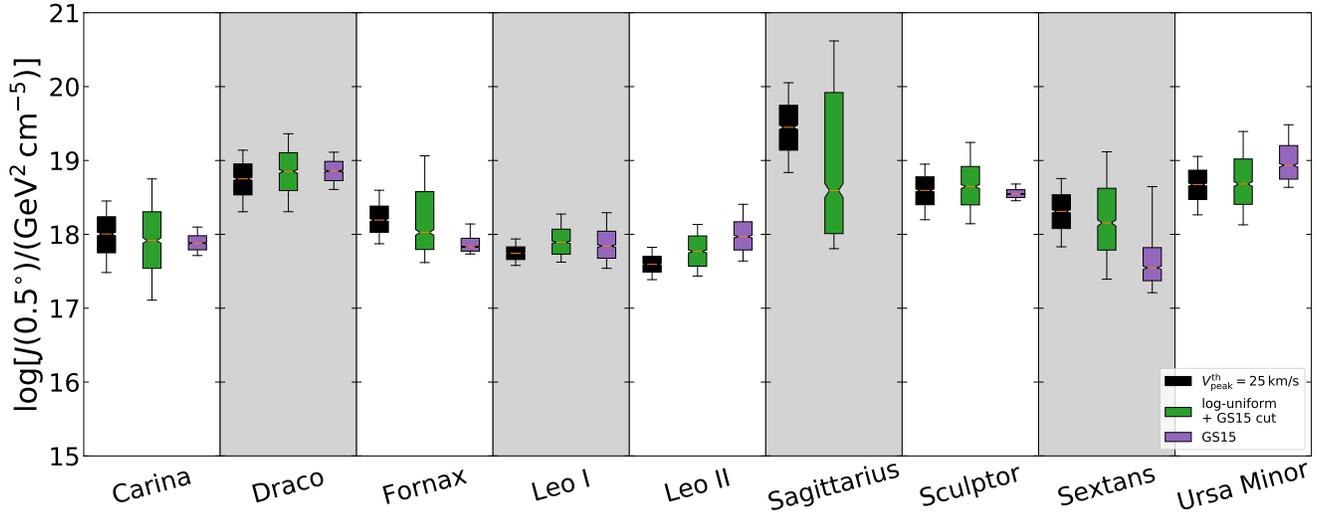

FIG. 4. Box and whisker plot of $J$ factors integrated within $0.5°$ for the classical dSphs. The boxes and whiskers show 68% and 95% credible equal-tailed intervals around the medians. The posteriors based on satellite priors with $V_{peak}^{th} = 25$ km s$^{-1}$ (black), log-uniform priors for ($\log r_s$, $\log \rho_s$) with GS15 cosmology cut (green), and the results from GS15 [45] are shown for comparison.

TABLE IV. The medians with 68% credible intervals of $\log[J(0.5°)/(\mathrm{GeV^2\ cm^{-5}})]$ for the ultrafaint dSphs. Second to the last columns correspond to different priors.

| Name | Log-uniform + GS15 cut | $V_{50} = 18$ km s$^{-1}$ | $V_{50} = 10.5$ km s$^{-1}$ | $V_{\mathrm{peak}}^{\mathrm{th}} = 6$ km s$^{-1}$ |
|---|---|---|---|---|
| Aquarius 2 | $17.92^{+0.59}_{-0.87}$ | $17.66^{+0.40}_{-0.47}$ | $17.21^{+0.52}_{-0.63}$ | $16.85^{+0.59}_{-0.56}$ |
| Bootes I | $17.87^{+0.37}_{-0.31}$ | $17.92^{+0.20}_{-0.22}$ | $17.80^{+0.23}_{-0.25}$ | $17.71^{+0.26}_{-0.29}$ |
| Bootes II | $19.86^{+0.96}_{-1.76}$ | $18.39^{+0.45}_{-0.53}$ | $17.69^{+0.66}_{-0.80}$ | $17.37^{+0.59}_{-0.62}$ |
| Canes Venatici I | $17.44^{+0.29}_{-0.13}$ | $17.39^{+0.12}_{-0.10}$ | $17.38^{+0.12}_{-0.11}$ | $17.38^{+0.12}_{-0.11}$ |
| Canes Venatici II | $17.84^{+0.47}_{-0.48}$ | $17.46^{+0.38}_{-0.39}$ | $17.21^{+0.40}_{-0.42}$ | $17.01^{+0.49}_{-0.51}$ |
| Carina II | $18.21^{+0.48}_{-0.61}$ | $18.29^{+0.32}_{-0.35}$ | $18.00^{+0.38}_{-0.45}$ | $17.77^{+0.42}_{-0.42}$ |
| Coma Berenices | $18.75^{+0.29}_{-0.35}$ | $18.55^{+0.25}_{-0.29}$ | $18.39^{+0.30}_{-0.34}$ | $18.29^{+0.34}_{-0.41}$ |
| Draco II | $19.30^{+1.03}_{-1.59}$ | $18.93^{+0.42}_{-0.47}$ | $18.31^{+0.60}_{-0.60}$ | $18.00^{+0.58}_{-0.58}$ |
| Eridanus II | $17.30^{+0.34}_{-0.35}$ | $16.94^{+0.23}_{-0.25}$ | $16.83^{+0.27}_{-0.30}$ | $16.80^{+0.28}_{-0.33}$ |
| Grus I | $17.31^{+1.65}_{-2.28}$ | $17.41^{+0.52}_{-0.56}$ | $16.61^{+0.71}_{-0.85}$ | $16.36^{+0.58}_{-0.61}$ |
| Hercules | $17.21^{+0.43}_{-0.50}$ | $17.15^{+0.29}_{-0.31}$ | $16.94^{+0.33}_{-0.38}$ | $16.75^{+0.36}_{-0.36}$ |
| Horologium 1 | $18.73^{+0.69}_{-0.90}$ | $18.00^{+0.42}_{-0.52}$ | $17.49^{+0.59}_{-0.68}$ | $17.10^{+0.65}_{-0.66}$ |
| Hyrdus 1 | $18.21^{+0.33}_{-0.36}$ | $18.27^{+0.22}_{-0.24}$ | $18.11^{+0.25}_{-0.29}$ | $18.00^{+0.26}_{-0.28}$ |
| Leo IV | $16.97^{+0.71}_{-1.15}$ | $17.06^{+0.43}_{-0.44}$ | $16.59^{+0.52}_{-0.64}$ | $16.33^{+0.49}_{-0.55}$ |
| Leo T | $17.43^{+0.41}_{-0.46}$ | $16.87^{+0.29}_{-0.35}$ | $16.66^{+0.38}_{-0.44}$ | $16.54^{+0.45}_{-0.49}$ |
| Leo V | $17.55^{+0.86}_{-1.16}$ | $17.17^{+0.47}_{-0.53}$ | $16.59^{+0.60}_{-0.71}$ | $16.25^{+0.59}_{-0.60}$ |
| Pegasus III | $17.93^{+0.88}_{-1.23}$ | $17.08^{+0.47}_{-0.55}$ | $16.46^{+0.64}_{-0.75}$ | $16.09^{+0.63}_{-0.62}$ |
| Pisces II | $17.97^{+0.87}_{-1.29}$ | $17.21^{+0.47}_{-0.55}$ | $16.57^{+0.65}_{-0.76}$ | $16.21^{+0.60}_{-0.62}$ |
| Reticulum II | $18.95^{+0.45}_{-0.48}$ | $18.73^{+0.32}_{-0.37}$ | $18.47^{+0.39}_{-0.45}$ | $18.23^{+0.46}_{-0.48}$ |
| Segue 1 | $19.56^{+0.38}_{-0.43}$ | $19.14^{+0.31}_{-0.36}$ | $18.92^{+0.38}_{-0.45}$ | $18.75^{+0.47}_{-0.55}$ |
| Segue 2 | $16.59^{+1.14}_{-1.93}$ | $17.94^{+0.41}_{-0.37}$ | $17.19^{+0.61}_{-0.84}$ | $17.16^{+0.49}_{-0.50}$ |
| Triangulum II | $17.39^{+1.57}_{-2.22}$ | $18.48^{+0.45}_{-0.50}$ | $17.67^{+0.68}_{-0.89}$ | $17.48^{+0.55}_{-0.58}$ |
| Tucana II | $18.94^{+0.62}_{-0.95}$ | $18.32^{+0.39}_{-0.48}$ | $17.80^{+0.57}_{-0.70}$ | $17.39^{+0.67}_{-0.60}$ |
| Tucana III | $15.42^{+1.44}_{-2.21}$ | $17.87^{+0.30}_{-0.30}$ | $16.72^{+0.70}_{-0.97}$ | $17.08^{+0.43}_{-0.36}$ |
| Ursa Major I | $18.49^{+0.25}_{-0.26}$ | $18.24^{+0.20}_{-0.25}$ | $18.18^{+0.22}_{-0.26}$ | $18.18^{+0.22}_{-0.22}$ |
| Ursa Major II | $19.34^{+0.34}_{-0.41}$ | $18.96^{+0.28}_{-0.32}$ | $18.78^{+0.35}_{-0.42}$ | $18.67^{+0.42}_{-0.55}$ |

## B. Ultrafaint dwarfs

In Figs. 5–10, we show the prior, likelihood, and posterior (68% and 95% credible or confidence regions) of $r_s$ and $\rho_s$ for all ultrafaint dSphs, with different model assumptions: $V_{50} = 18$ km s$^{-1}$ and 10.5 km s$^{-1}$, and $V_{\mathrm{peak}}^{\mathrm{th}} = 6$ km s$^{-1}$.

As above, we compute the astrophysical $J$ factors within a radius of $0.5°$ and their posterior distributions based on those of $(r_s, \rho_s, r_t)$, adopting approximate formulae given in Ref. [46]. In Figs. 11–13, we show the posteriors for $J$ for satellite priors with $V_{50} = 18$ km s$^{-1}$ and 10.5 km s$^{-1}$ as well as uniform priors of $(\log r_s, \log \rho_s)$ with and without GS15 cut [45] corresponding to the gray shaded region of Figs. 5–10. The medians and 68% and 95% credible intervals of $\log[J/(\mathrm{GeV^2\ cm^{-5}})]$ for the ultrafaints are shown in Fig. 3 of the main text and in Table IV.

## V. EFFECT OF THE SATELLITE PRIORS FOR OTHER ANNIHILATION CHANNELS

While we only consider the annihilation of dark matter via the $b\bar{b}$ channel in the main text, it is important to note that the satellite priors will affect any other annihilation channel or dark matter model in a similar way. The choice of dark matter model only affects the cross section, while the $J$ factor can be determined independently. Therefore, the impact of the satellite prior will be approximately the same order of magnitude across different annihilation channels given a particular analysis. However, the exact numerical values of limits will depend on the spectral information contained in the data and the predicted spectral shape for that particular annihilation channel. To illustrate this, we show the limits for the $\tau^+\tau^-$ channel in Fig. 14. We observe a qualitatively similar picture to the case for $b\bar{b}$ presented in the main text.

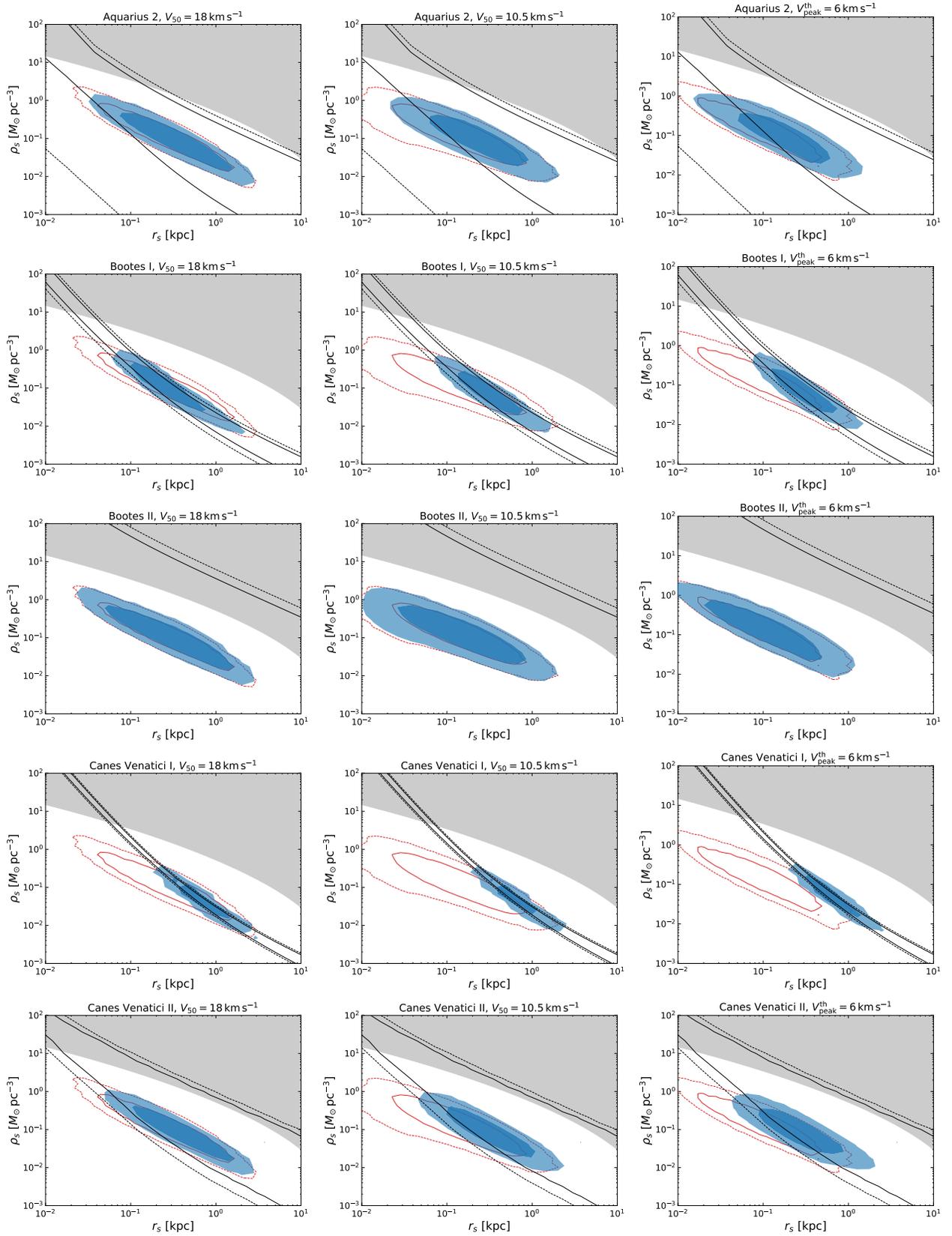

FIG. 5. Satellite priors (red), likelihoods (black), and posteriors (filled blue) of $(r_s, \rho_s)$ for ultrafaint dSphs analyzed in this work at 68% and 95% credibility or confidence levels. The priors correspond to the satellite forming condition of $V_{50} = 18$ km s$^{-1}$ (left), $V_{50} = 10.5$ km s$^{-1}$ (middle), and $V_{peak}^{rth} = 6$ km s$^{-1}$ (right). The gray shaded region shows the cosmology cut implemented in Ref. [45] for the log-uniform priors. For dSphs like Bootes II, which have unpaired likelihood contours, the lower limits on $\rho_s$ are off the bottom of the plot. In these cases the measured velocity dispersion is consistent with zero.

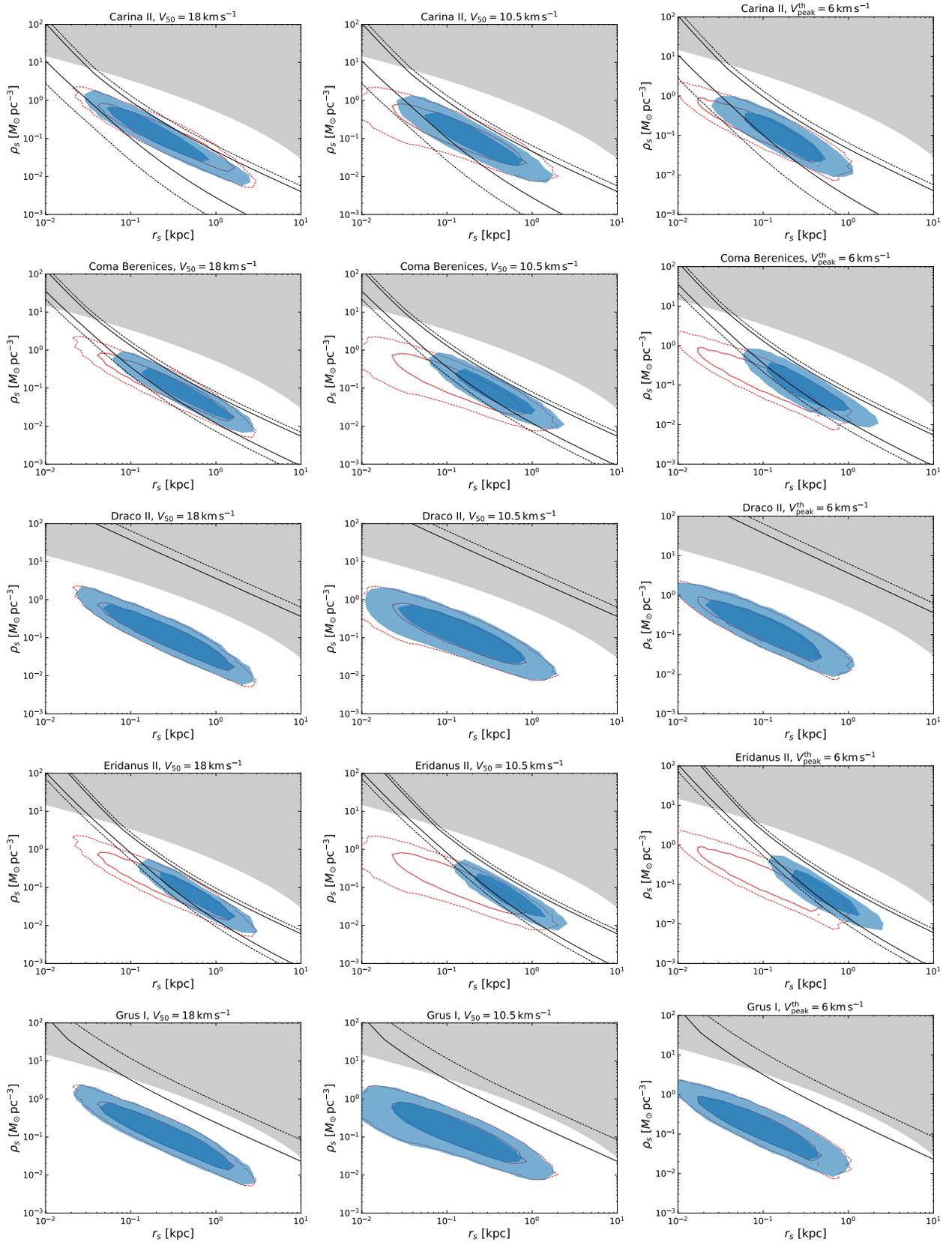

FIG. 6. Continued from Fig. 5.

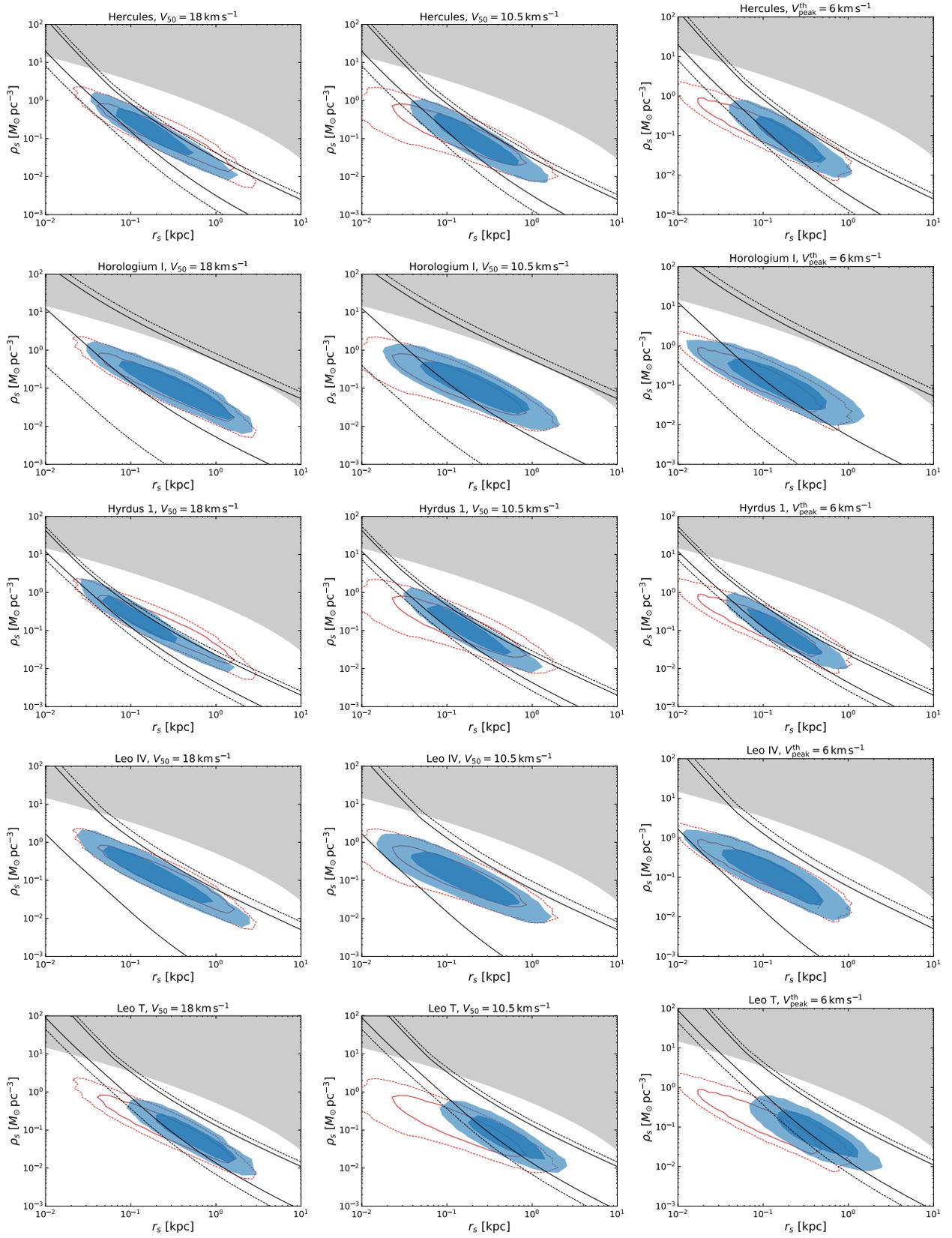

FIG. 7. Continued from Fig. 6.

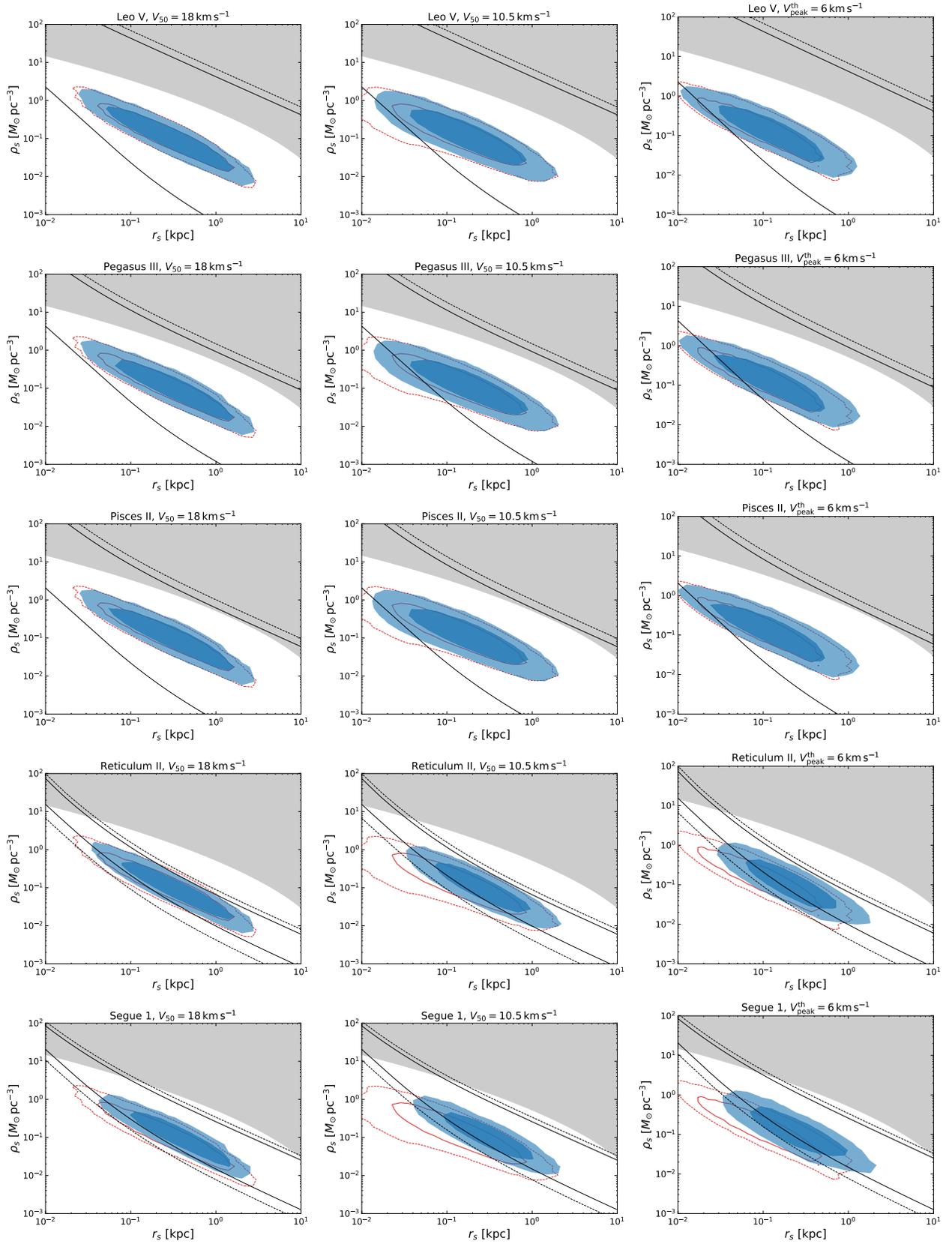

FIG. 8. Continued from Fig. 7.

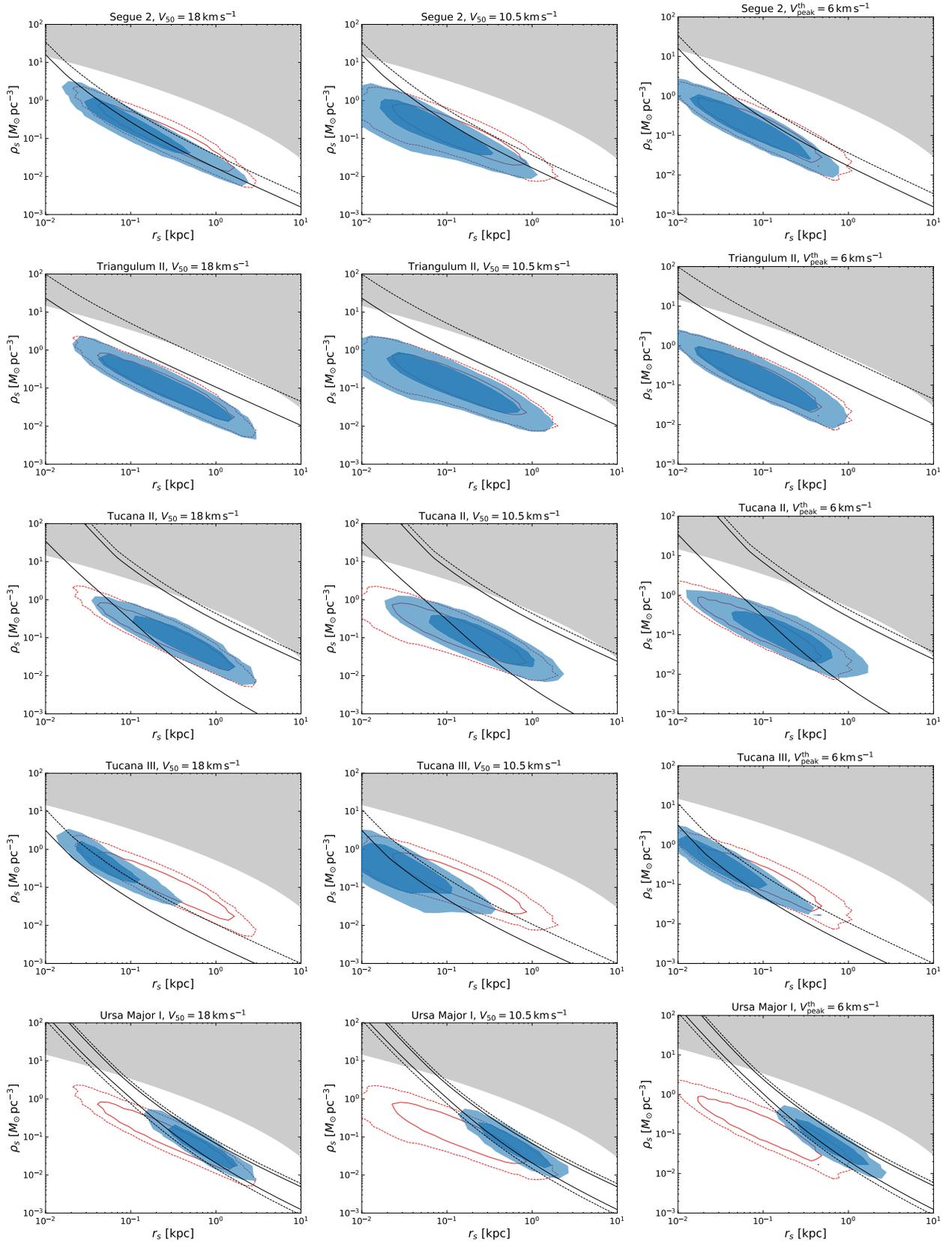

FIG. 9. Continued from Fig. 8.

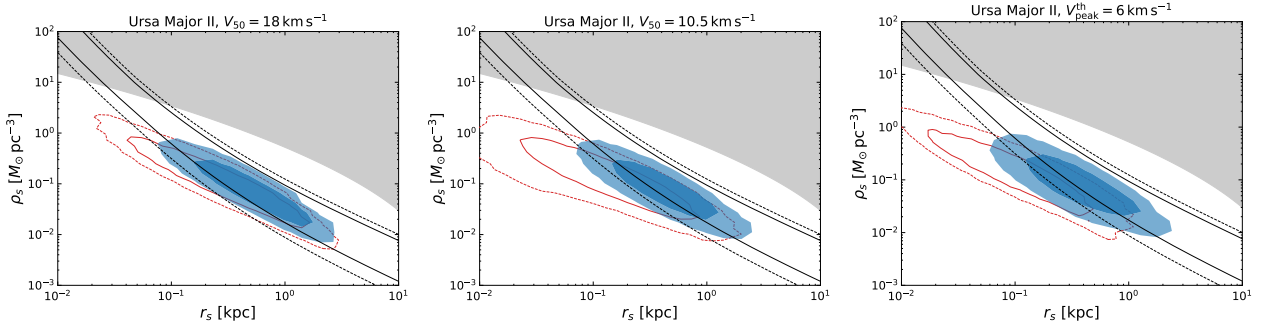

FIG. 10. Continued from Fig. 9.

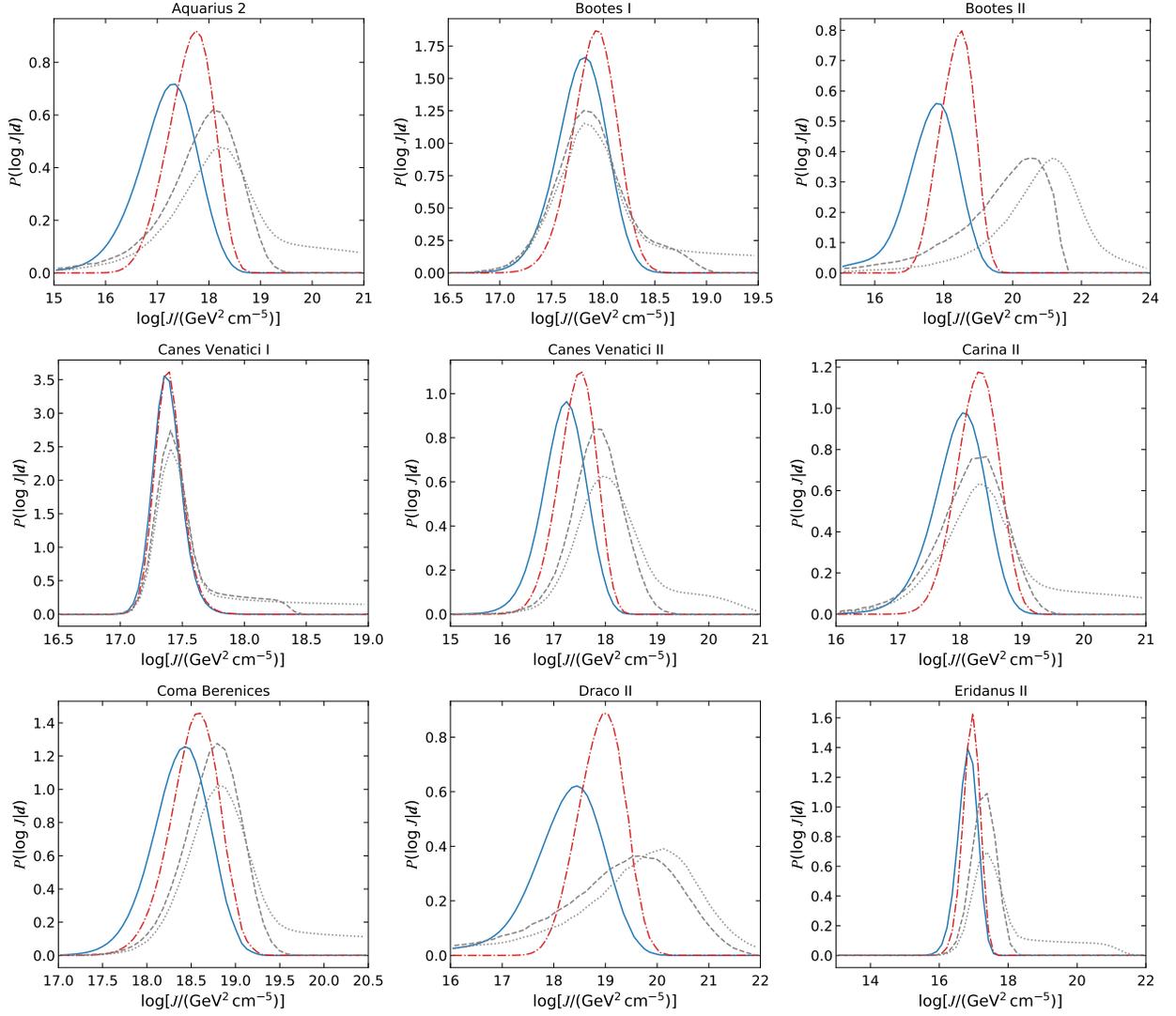

FIG. 11. Posterior distributions of $\log[J(0.5°)/(\text{GeV}^2\,\text{cm}^{-5})]$ for ultrafaint dSphs obtained with satellite priors with $V_{50} = 10.5$ (blue solid) and 18 km s$^{-1}$ (red dot-dashed), log-uniform priors with (gray dashed) and without (gray dotted) the GS15 cut [45].

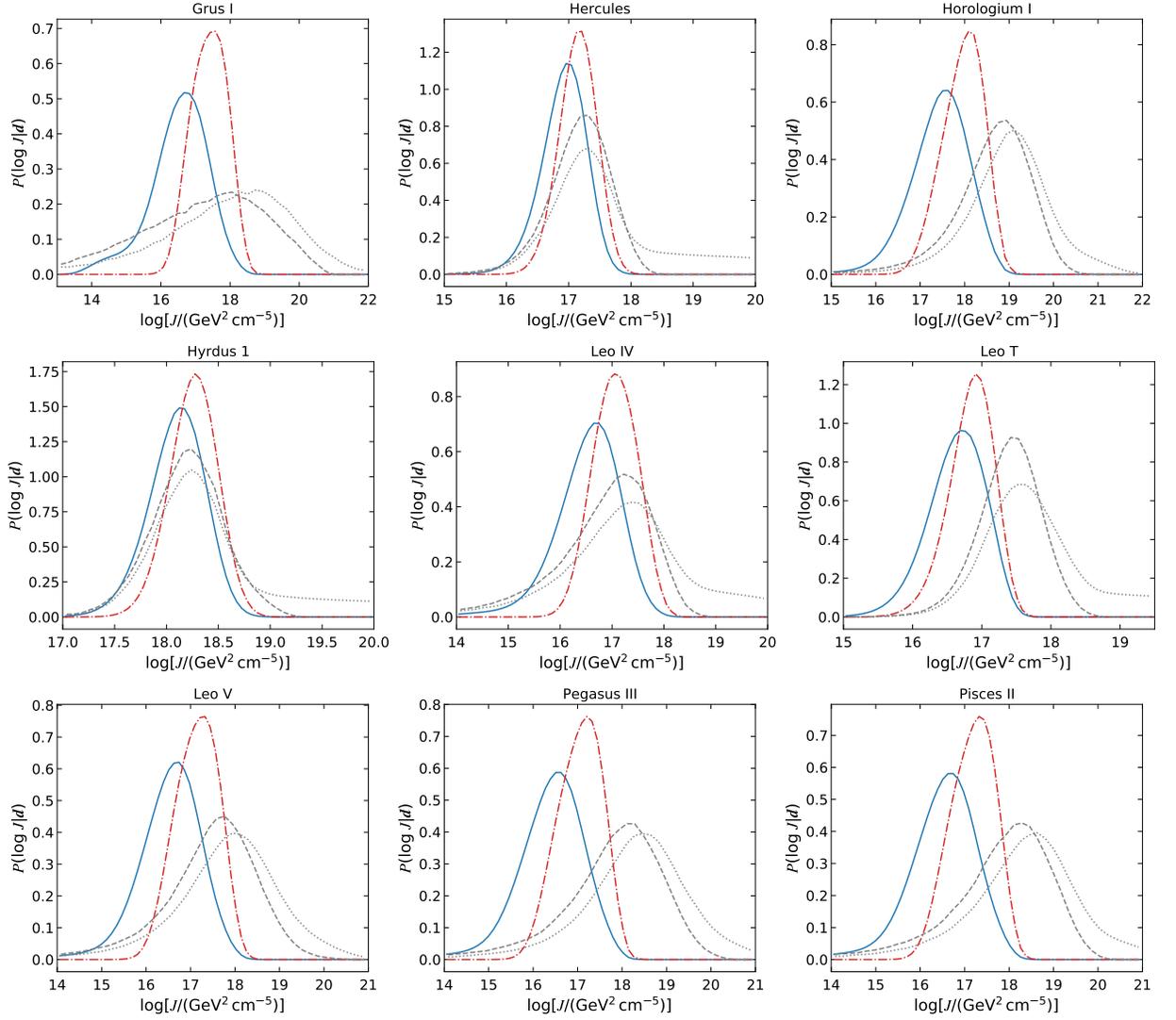

FIG. 12. Continued from Fig. 11.

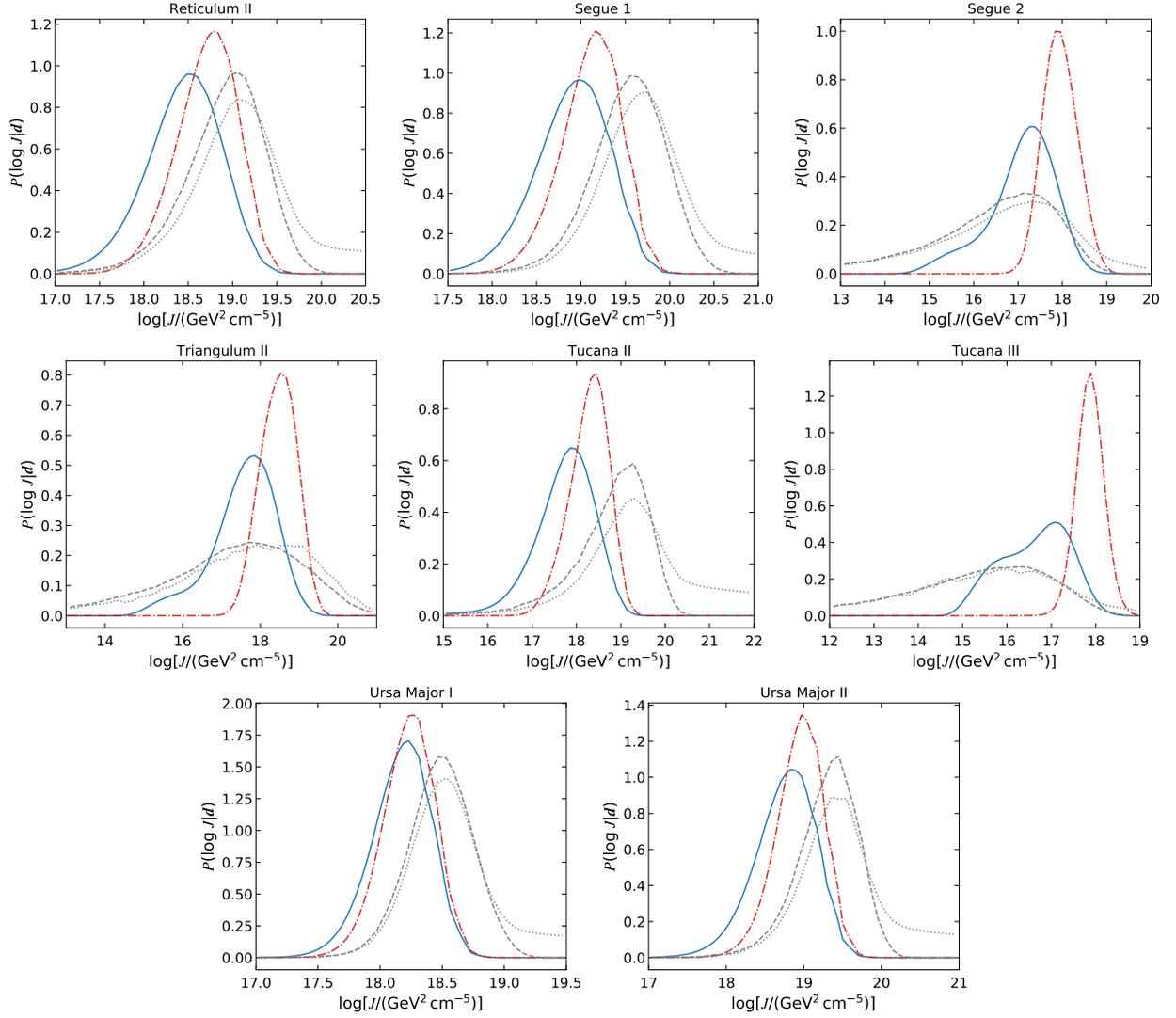

FIG. 13. Continued from Fig. 12.

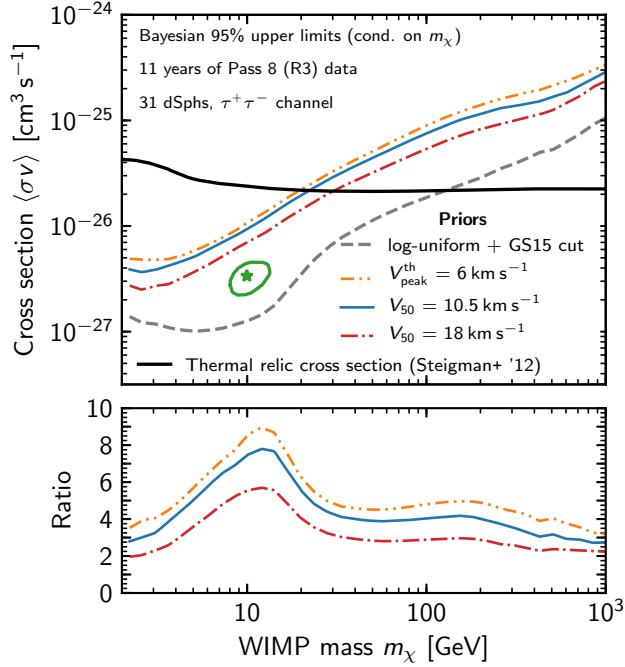

FIG. 14. Limits on the WIMP annihilation cross section $\langle \sigma v \rangle$ ($\tau^+\tau^-$ channel) for different prior choices. *Top:* Limits at 95% credibility (conditioned on the WIMP mass $m_\chi$). The star and surrounding region indicate the parameter point and $2\sigma$ confidence region associated with a possible Galactic centre excess [47]. *Bottom:* Ratios of limits based on the log-uniform priors with GS15 cut to those obtained with satellite priors, i.e., how much *weaker* the limits derived from the satellite priors are.